\documentclass[12pt]{article}

\usepackage{amssymb}
\usepackage{amsmath}
\usepackage{amscd}
\usepackage{latexsym}
\usepackage{graphicx}

\usepackage{cite}

\newcommand{\be}{\begin{equation}}
\newcommand{\ee}{\end{equation}}

\newcommand{\dlt}{\delta}

\newcommand{\ep}{\varepsilon}

\newcommand{\ra}{\rightarrow}

\begin{document}

\begin{center}

{\Large{\bf Self-excited waves in complex social systems} \\ [5mm]

V.I. Yukalov$^{1,2,*}$, E.P. Yukalova$^{3}$ } \\ [3mm]

{\it
$^1$Bogolubov Laboratory of Theoretical Physics, \\
Joint Institute for Nuclear Research, Dubna 141980, Russia \\ [3mm]

and \\

Instituto de Fisica de S\~ao Carlos, Universidade de S\~ao Paulo, \\
CP 369,  S\~ao Carlos 13560-970, S\~ao Paulo, Brazil  \\ [2mm]
                          
$^3$Laboratory of Information Technologies, \\
Joint Institute for Nuclear Research, Dubna 141980, Russia} \\ [3mm]

$^*${\it Corresponding author e-mail}: yukalov@theor.jinr.ru \\ [3mm]

\end{center}

\vskip 2cm

\begin{abstract}
A social system is considered whose agents choose between several alternatives of 
possible actions. The system is described by the fractions of agents preferring
the corresponding alternatives. The agents interact with each other by exchanging 
information on their choices. Each alternative is characterized by three attributes:
utility, attractiveness, and replication. The agents are heterogeneous having
different initial conditions and different types of memory, which can be long-term
or short-term. The agent interactions, generally, can depend on the distance between
the agents, varying from short-range to long-range interactions. The emphasis in the
paper is on long-range interactions. In a mixed society consisting of agents with 
both long-term and short-term memory, there appears the effect of spontaneous
excitation of preference waves, when the fractions of agents preferring this or that
alternative suddenly start strongly oscillating, either periodically or chaotically. 
Since the considered society forms a closed system, without any external influence, 
the arising waves are internally self-excited by the society. 
\end{abstract}

\vskip 1cm

{\parindent=0pt

{\bf Keywords}: complex social systems, information exchange, long-term memory, 
short-term memory, herding effect, spontaneous preference waves 
}

\newpage

\section{Introduction}

Understanding the behaviour of social systems is of long standing and permanent 
interest \cite{Parsons_1}, and it has been described employing various models 
based on statistical physics (see, e.g., reviews \cite{Perc_2,Perc_3}) or on the 
theory of networks \cite{Scott_4,Albert_5,Boccaletti_6,Meter_7,Kennett_8,Mata_9}. 
The recent review \cite{Jusup_3} summarizes both approaches. Here we keep in mind 
social networks consisting of agents choosing between different alternatives. 
Interactions between society members usually are described by given functions, 
similarly to the interactions between particles or spins in statistical mechanics. 
In that sense, society members are treated as nodes mechanically accomplishing 
prescribed actions. In the case of realistic biological, especially human, societies, 
such models, clearly, give a rather simplified picture, since the agents of these 
societies are not mechanical devices, but are intelligent agents. The basic feature 
of an intelligent agent is the ability, after evaluating the available information, 
to make decisions choosing between several alternative actions 
\cite{Nilsson_10,Poole_11,Luger_12,Rich_13,Russell_14}.  

The aim of the present paper, is to suggest a model of a society composed of 
intelligent agents. To our understanding, such a model could provide a more 
realistic description of social systems consisting of intelligent agents taking 
decisions with respect to available alternatives. The basic points of the suggested 
theory are as follows. 

(i) The approach is probabilistic. Each agent is characterized by the probability 
of choosing this or that alternative. This probability has the meaning of a 
frequentist measure showing the fraction of agents choosing the given alternative.

(ii) The probability measure takes into account three factors: the utility of 
alternatives, their attractiveness, and the attitude of agents towards replicating
the actions of other members of the society. Thus, in addition to the estimation 
of utility of alternatives, our model includes the influence of emotions, described
by the attractiveness of alternatives, and takes into account the herding effect.

Endeavors of including emotions in the process of choice have been undertaken in the 
frame of quantum decision theory 
\cite{Yukalov_15,Yukalov_16,Yukalov_17,Yukalov_18,Yukalov_19,Yukalov_20,Yukalov_21,Yukalov_YYS}.
However, as has been shown \cite{Yukalov_22}, the use of quantum theory can be avoided,
and an approach can be developed incorporating emotions into decision process, without
resorting to quantum techniques. Below we suggest an approach of describing the society
of intelligent agents, employing only classical notions. We study a novel kind of
complex society composed of agents with different types of memory, so that a fraction
of the society members is endowed with long-term memory, while the others have 
short-term memory. In this complex society, interesting new effects appear, such as 
suddenly arising self-excited oscillations of preferences. These spontaneous oscillations
can be either periodic or chaotic.

The layout of the paper is as follows. In Sec. 2, we formulate the model of a society
of intelligent agents deciding between several alternatives. The choice is based of three
attributes, utility, attractiveness, and herding. In Sec. 3, the problem is specified
to the consideration of two groups of agents choosing between two alternatives. The groups
are composed of agents possessing different types of memory, either long-term or short-term
memory. Section 4 presents a detailed investigation of different dynamic regimes that can
occur in the process of decision making. Section 5 concludes.

\section{Society of intelligent agents}

Consider a society of $N$ agents, enumerated by the index $j=1,2,\ldots,N$. Assume that
each agent needs to choose between $N_A$ alternative actions $A_n$, numbered by the 
index $n=1,2,\ldots,N_A$. This can be the choice between several candidates at elections, 
between different goods in a shop, between several jobs, etc. 

The probability that a $j$-th agent chooses an alternative $A_n$ at time $t$ is 
denoted as $p_j(A_n,t)$. By the meaning of probability, $p_j(A_n,t)$ has to be 
non-negative and normalized,
\be
\label{1}
\sum_{n=1}^{N_A} p_j(A_n,t) = 1 \; , \qquad 0 \leq p_j(A_n,t) \leq 1 \; .
\ee
         
The utility of an alternative $A_n$ defines the weight $f(A_n,t)$ ascribed to this 
alternative, because of which $f(A_n,t)$ can be called {\it utility factor}. Being 
a  weight, it enjoys the properties of classical probability,
\be
\label{2}
\sum_{n=1}^{N_A} f_j(A_n,t) = 1 \; , \qquad 0 \leq f_j(A_n,t) \leq 1 \;   .
\ee
The explicit expression for the utility factor can be found from the minimization of 
an information functional \cite{Yukalov_19,Yukalov_23}.

Important role in the process of decision making is played by emotions that can be 
characterized by the {\it attraction factor} $q(A_n,t)$. Emotions can be positive 
or negative, because of which the attraction factor varies in the interval $[-1,1]$, 
\be
\label{3}
-1 \leq q_j(A_n,t) \leq 1 \;  .
\ee

If the agents of a society exchange information with each other, there appears herding
effect when the members of the society incline to replicate the actions of others. Let
this effect be quantified by the {\it herding factor} $h(A_n,t)$, being in the interval
\be
\label{4}
-1 \leq h_j(A_n,t) \leq 1 \; .
\ee

The probability of preferring an alternative $A_n$ is the superposition of the utility,
attraction, and herding factors. Taking into account the fact that the process of making 
a decision requires some time, say $\tau$, we have
\be
\label{5}
 p_j(A_n,t+\tau) = f_j(A_n,t) + q_j(A_n,t) + h_j(A_n,t) \; .
\ee
This expression differentiates the probabilistic approach we follow from multi-attribute 
utility theory, where the expected utility functional is represented as a superposition 
of weighted parts associated with different attributes \cite{Fishburn_24,Keeney_25}.  

From the normalization conditions (\ref{1}) and (\ref{2}), it follows that the sum of 
the attraction factor and herding factor over all alternatives is zero:
\be
\label{6}
\sum_n \left[ \;  q_j(A_n,t) + h_j(A_n,t) \; \right] = 0 \; .
\ee
Assuming that attraction and herding are independent, we have
\be
\label{7}
\sum_n   q_j(A_n,t) = 0 \; , \qquad  \sum_n  h_j(A_n,t)  = 0 \; .
\ee
The latter equations can be called {\it alternation law}.  
   
Although, on a long time scale there exists time discounting \cite{Frederick_26}, 
however the utility of alternatives varies in time slowly, which allows us to consider 
it constant,
\be
\label{8}
  f_j(A_n,t) =  f_j(A_n) \; .
\ee

On the contrary, information exchange between the society agents is a fast process, 
such that the attraction factor of a $j$-th agent essentially depends on the amount 
of information $M_j(t)$ obtained by this agent. The attraction factor can be modeled
\cite{Yukalov_27} by the expression
\be
\label{9}
 q_j(A_n,t) = q_j(A_n) \exp\{ - M_j(t) \} \;  ,
\ee 
where $q(A_n) = q(A_n,0)$ is an initial value of the attraction factor. This 
expression is similar to the time dependence of decoherence factor under nondestructive 
repeated measurements \cite{Yukalov_28,Yukalov_29}. The amount of information obtained 
by a $j$-th agent by the time $t$ can be written as
\be
\label{10}
M_j(t) = \sum_{t'=0}^t \; \sum_{i=1}^N J_{ji}(t,t') \; \mu_{ji}(t') \;  .
\ee
This can be called {\it remembered information}. Here $J_{ji}(t,t')$ is the intensity 
of interactions during the information transfer from an $i$-th agent to the $j$-th 
agent in the period of time between $t'$ and $t$. The information gain, received by 
the $j$-th agent from an $i$-th agent at time $t$, is given in the form of the 
Kullback-Leibler \cite{Kullback_30,Kullback_31} relative information
\be
\label{11}
 \mu_{ji}(t) = 
\sum_{n=1}^{N_A} p_j(A_n,t) \; \ln \; \frac{p_j(A_n,t)}{p_i(A_n,t)} \; .
\ee
Note that the information gain enjoys the properties $\mu_{ji} \geq 0$ and 
$\mu_{jj} = 0$. At the initial moment of time, no information has yet been transferred, 
implying that
\be
\label{12}
M_j(0) = 0 \;  .
\ee

The interactions $J_{ij}(t,t')$, in general, depend on the distance between the 
agents. In the case of short-range interactions the topology of the agent locations 
is important. In the opposite case of long-range interactions, the geometry of the 
agent network plays no role, with the interactions acquiring the form 
\be
\label{13}
 J_{ij}(t,t') = \frac{1}{N-1} \;  J(t,t') \; .
\ee
Our concern here is a society composed of intelligent agents, like humans who are 
able to interact with each other irrespectively of the distance between them. This 
kind of distance-independent interactions are provided nowadays by internet, mass 
media, and phones. In addition, the agents are not located at fixed nodes but can 
freely move varying their whereabouts. Keeping in mind these location-independent 
interactions, we accept in what follows the form (\ref{13}). Then the amount of 
information kept in the agent memory reads as
\be
\label{14}
M_j(t) = 
\sum_{t'=0}^t \frac{J(t,t')}{N-1} \; \sum_{i=1}^{N} \mu_{ji}(t') \; .
\ee

The agent memory is characterized by its duration. In two opposite situations, it 
can be long-term or short-term. The ultimate long-term memory is permanent in time,
\be
\label{15}
J(t,t') = J \qquad ( long - term)  \;  .
\ee
Then the amount of remembered information (\ref{14}) takes the form
\be
\label{16}
M_j(t) =  \frac{J}{N-1} \; 
\sum_{t'=0}^t \; \sum_{i=1}^{N} \mu_{ji}(t') \; .
\ee

The opposite case is the short-term memory, when the information from only the last 
step is remembered,
\be
\label{17}
J(t,t') = J \;\dlt_{tt'} \qquad ( short - term) \;  .
\ee
In that case, the available remembered information (\ref{14}) becomes
\be
\label{18}
M_j(t) = \frac{J}{N-1} \sum_{i=1}^{N} \mu_{ji}(t) \; .
\ee
 
In complex societies, there exists a collective effect, when the agents are prone to
imitate the actions of others. This is called the herd effect. Herd behavior occurs 
in animal herds, packs, bird flocks, fish schools and so on, as well as in humans. 
It is well known and studied for many years 
\cite{Martin_32,Sherif_33,Smelser_34,Merton_35,Turner_36,Hatfield_37,Brunnermeier_38}.   
In evolution equations of social and biological systems, the mathematical description 
of herding is represented by the replication term \cite{Sandholm_39}, which for our 
case takes the form
\be
\label{19}
 h_j(A_n,t) = \ep_j \; \left\{ 
\frac{1}{N-1} \sum_{i(\neq j) }^N [\; f_i(A_n) + q_i(A_n,t) \; ] -
 [\; f_j(A_n) + q_j(A_n,t) \; ] \right\} \;  .
\ee
The parameters $\ep_j$ describe the intensity of the herding behaviour. Due to the 
normalization conditions (\ref{1}), (\ref{2}), (\ref{3}), (\ref{4}), and (\ref{7}), 
these parameters satisfy the inequalities
\be
\label{20}
 0 \leq \ep_j \leq 1 \qquad ( j = 1,2,\ldots, N) \; .
\ee

Without the loss of generality, time can be measured in units of $\tau$. Substituting
the herding term (\ref{19}) into expression (\ref{5}) yields the equation 
\be
\label{21}
 p_j(A_n,t+1) = (1- \ep_j ) \; [\; f_j(A_n) + q_j(A_n,t) \; ] + 
\frac{\ep_j}{N-1} \sum_{i(\neq j)}^N [\; f_i(A_n) + q_i(A_n,t) \; ] \; .
\ee
This it the evolution equation for the probability that the $j$-th agent chooses at
time $t+1$ the alternative $A_n$. The right-hand side of this equation at time $t=0$
gives the initial condition
\be
\label{22}
p_j(A_n,0) = (1- \ep_j ) \; [\; f_j(A_n) + q_j(A_n,0) \; ] + 
\frac{\ep_j}{N-1} \sum_{i(\neq j)}^N [\; f_i(A_n) + q_i(A_n,0) \; ] \;  .
\ee

\section{Agents with different types of memory}

Now we need to specify what kind of agents we consider. Agents can differ by initial 
conditions or by the types of their memory. The simplest case is when all agents 
possess the same type of memory, although even then the dynamics of preferences can 
exhibit rather nontrivial behaviour \cite{Yukalov_27}. Here we study a more interesting 
case, where the society consists of agents with different types of memory. A fraction 
of agents has long-term memory and the other part, short-term memory. Such a mixture 
of agents with different properties much better models the real human societies. It 
turns out that this more complex society exhibits unusual phenomena that are absent in 
homogeneous societies where all agents possess the same type of memory. For instance,
there appear self-excited waves of preferences, spontaneously arising without any 
external influence. These waves can appear at the beginning of dynamics or may not 
exist at the beginning of decision processes, but appear suddenly after some time of 
quite smooth dynamics.      

Suppose the society consists of two types of agents. One part enjoys long-term memory
associated with the information function (\ref{16}), while the other part has short-term 
memory corresponding to the information function (\ref{18}). A group of similar agents
can be represented by a frequentist probability showing the fraction of agents choosing
an alternative $A_n$ at time $t$. For the case of two groups, we have the probabilities
$p_1(A_n,t)$ and $p_2(A_n,t)$. 

Let us also consider the very often met situation when the choice is between two 
alternatives, say $A_1$ and $A_2$. Then, keeping in mind the normalization conditions 
(\ref{1}), (\ref{2}), and (\ref{7}), with $j = 1,2$, we can simplify the notation for
the probabilities,
\be
\label{23}     
p_j(A_1,t) \equiv p_j(t) \; , \qquad p_j(A_2,t) = 1 - p_j(t) \;   ,
\ee
utility factors,
\be
\label{24}
f_j(A_1) \equiv f_j \; , \qquad f_j(A_2) = 1 - f_j \; ,
\ee
attraction factors,
\be
\label{25}
 q_j(A_1,t)\equiv q_j(t) \; , \qquad q_j(A_2,t) =  - q_j(t) \;   ,
\ee
and herding factors,
\be
\label{26}
 h_j(A_1,t) \equiv h_j(t) \; , \qquad h_j(A_2,t) =  - h_j(t) \;   .
\ee
Then we have the probabilities
\be
\label{27}
p_j(t+1) = f_j + q_j(t) + h_j(t) \qquad ( j = 1,2 ) \; ,
\ee
with the attraction and herding factors 
$$
q_j(t) = q_j \exp\{ - M_j(t) \} \qquad ( j = 1,2 ) \; ,
$$
\be
\label{28}
 h_j(t) = \ep_j \; [\; f_i + q_i(t) - f_j - q_j(t) \; ] \qquad 
( i\neq j ) \; ,
\ee
where $q_j \equiv q_j(0)$. 

Let us mark the group of agents with long-term memory as the group number one, and 
the group of agents with short-term memory, as the second group. For the corresponding 
available remembered information, setting $J =1$, we have in the case of long-term 
memory 
\be
\label{29}
  M_1(t) = \sum_{t'=0}^t \mu_{12}(t') 
\ee
and for short-term memory, 
\be
\label{30}
  M_2(t) = \mu_{21}(t) \;  .
\ee
The information gain reads as
\be
\label{31}
 \mu_{ij}(t)= p_i(t) \; \ln \; \frac{p_i(t)}{p_j(t)} + 
[\; 1 - p_i(t) \;] \; \ln \; \frac{1-p_i(t)}{1-p_j(t)} \; ,
\ee
where $i \neq j$. Thus we come to the evolution equations
$$
 p_1(t+1) = 
( 1 - \ep_1)\; [\; f_1 + q_1(t) \; ] + \ep_1 \; [\; f_2 + q_2(t) \; ] \; ,
$$ 
\be
\label{32}
p_2(t+1) = 
( 1 - \ep_2)\; [\; f_2 + q_2(t) \; ] + \ep_2 \; [\; f_1 + q_1(t) \; ] \; ,
\ee
with the initial conditions
$$
 p_1(0) = 
( 1 - \ep_1)\; [\; f_1 + q_1 \; ] + \ep_1 \; [\; f_2 + q_2 \; ] \; ,
$$
\be
\label{33}
 p_2(0) = 
( 1 - \ep_2)\; [\; f_2 + q_2 \; ] + \ep_2 \; [\; f_1 + q_1 \; ] \;  
\ee
that are defined by the parameters $f_1$, $q_1$, $f_2$, and $q_2$. 

Recall that $p_1(t)$ is the fraction of agents with long-term memory preferring the 
alternative $A_1$ at time $t$, while $p_2(t)$ is the fraction of agents with short-term
memory preferring the alternative $A_1$ at time $t$. Because of the exchange of 
information and the herding effect, these fractions vary in time.

\section{Dynamics of preferences}

Before going to the numerical investigation of Eqs. (\ref{32}), it is possible to make
general conclusions on the expected role of the herding effect. To this end, it is easy 
to notice that, when the herding behavior is absent, hence $\ep_1=\ep_2=0$, then 
Eqs. (\ref{32}) read as
$$
p_1(t+1) = f_1 + q_1(t) \qquad ( \ep_1 = 0 ) \; ,
$$
\be
\label{34}
p_2(t+1) = f_2 + q_2(t) \qquad ( \ep_2 = 0 ) \;   .
\ee
This does not mean that the first and second groups are independent, since they remain 
connected through the information exchange entering the remembered information (\ref{29})
and (\ref{30}). However, to some extent, the evolution of group preferences can be quite
different because of so different memory properties of the groups. 

Increasing the herding parameters $\ep_j > 0$ switches on the herding behavior that, 
due to the replication terms $h_j(t)$, should smooth the differences in the dynamics 
of the probabilities $p_j(t)$. When the herding parameters reach the value
$\ep_1=\ep_2=1/2$, the behaviour of both groups becomes identical, since
\be
\label{35}
p_1(t+1) = p_2(t+1)  \qquad \left( \ep_1 = \ep_2 = \frac{1}{2} \right) \;  .
\ee

With the following increase of the herding parameters, the difference in the behaviour 
of the groups starts growing, reaching the maximum for $\ep_1=\ep_2=1$, when the 
probabilities become
$$
 p_1(t+1) = f_2 + q_2(t) \qquad ( \ep_1 = 1) \; ,
$$
\be
\label{36}
 p_2(t+1) = f_1 + q_1(t) \qquad ( \ep_2 = 1) \;  .
\ee

Thus the expected behaviour of the group probabilities can be rather different at the 
point $\ep_1+\ep_2=0$, when there is no herding effect. The difference smooths when 
approaching the line $\ep_1+\ep_2 = 1$. An then the difference increases again when 
drifting to the line $\ep_1+\ep_2 = 2$. Numerical solution of Eqs. (\ref{32}) 
demonstrates that there occur various types of qualitatively different behaviour. In 
the first four Figures 1 to 4, we show dynamic regimes corresponding to what can be 
called moderate herding, with the herding parameters varying from the point 
$\ep_1+\ep_2=0$ to the line $\ep_1+\ep_2=1$. Figures 5 to 10 present dynamic regimes 
for large herding parameters between the lines $\ep_1+\ep_2=1$  and $\ep_1+\ep_2=2$. 
Overall, the following dynamic regimes can happen.  

\vskip 2mm
{\bf 1}. In the absence of herding, the functions $p_j(t)$ tend to fixed points
\be
\label{37}
p_j^* = \lim_{t\ra\infty} p_j(t) \;  .
\ee
The appearance of herding makes the trajectories closer to each other. If there are 
sharp variations of the functions, they become smoothed by the herding effect, as is 
shown in Fig. 1.

%Figure 1
\begin{figure}[ht]
\centerline{
\hbox{ \includegraphics[width=7.5cm]{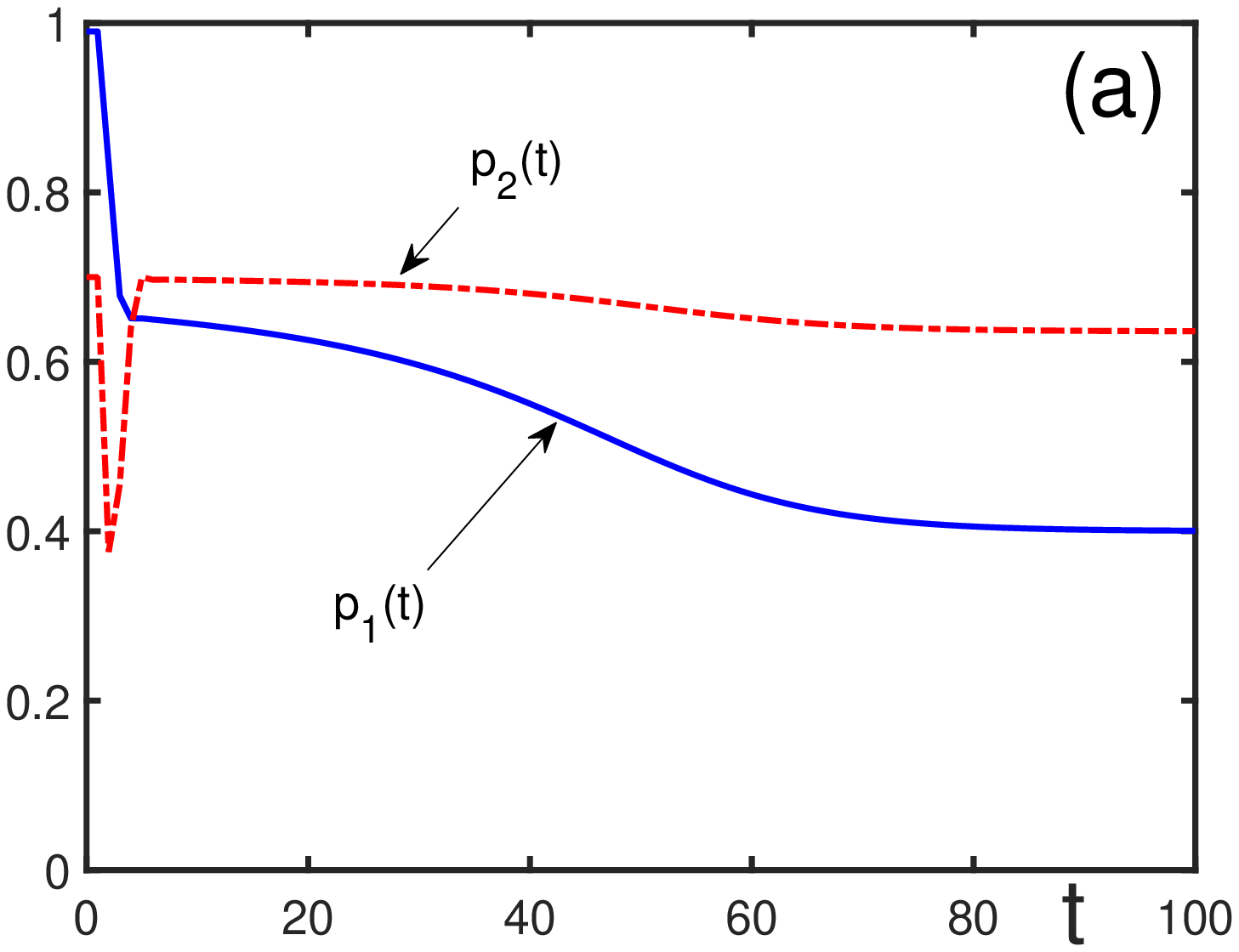} \hspace{1cm}
\includegraphics[width=7.5cm]{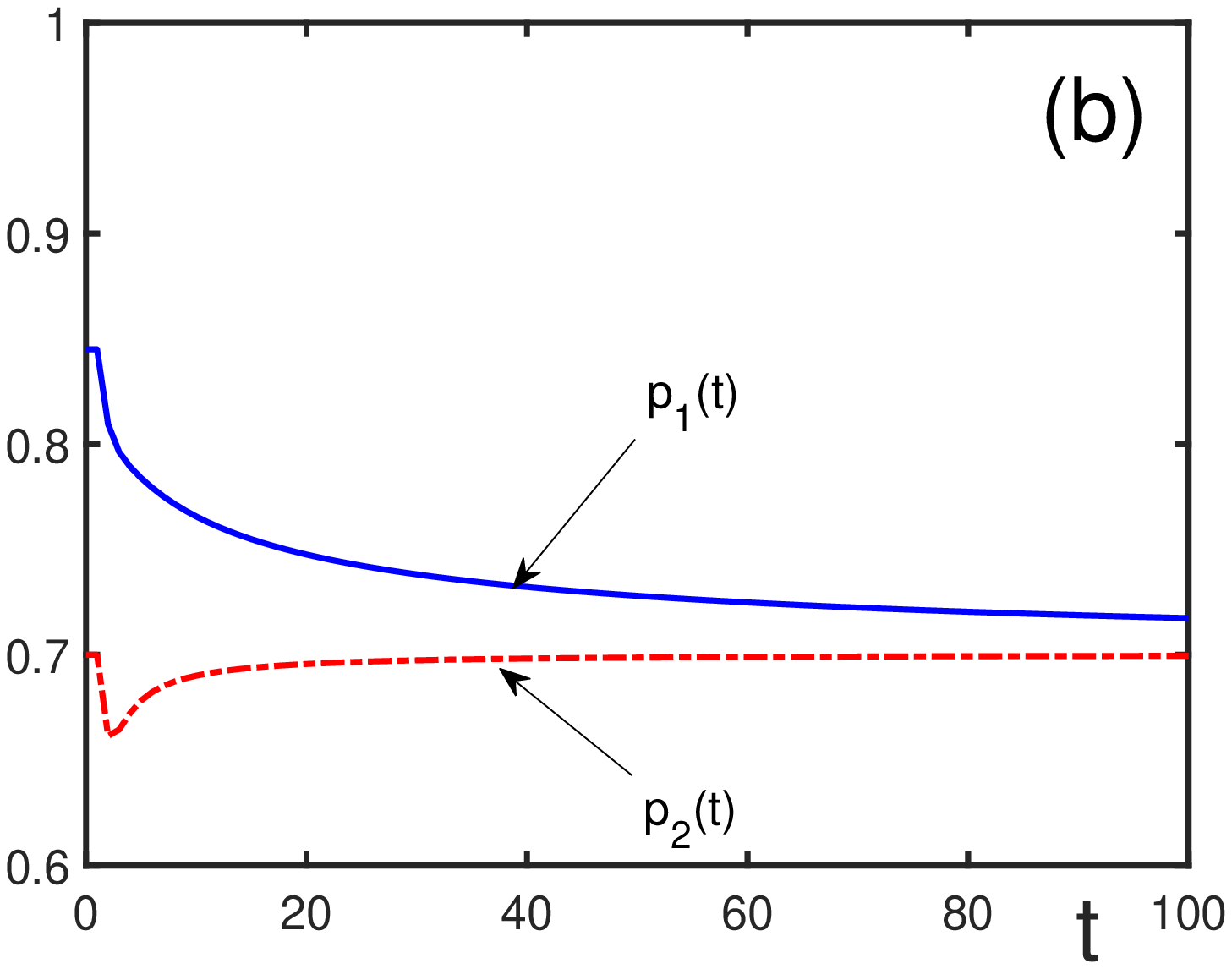}  } }
\vspace{12pt}
\centerline{
\hbox{ \includegraphics[width=7.5cm]{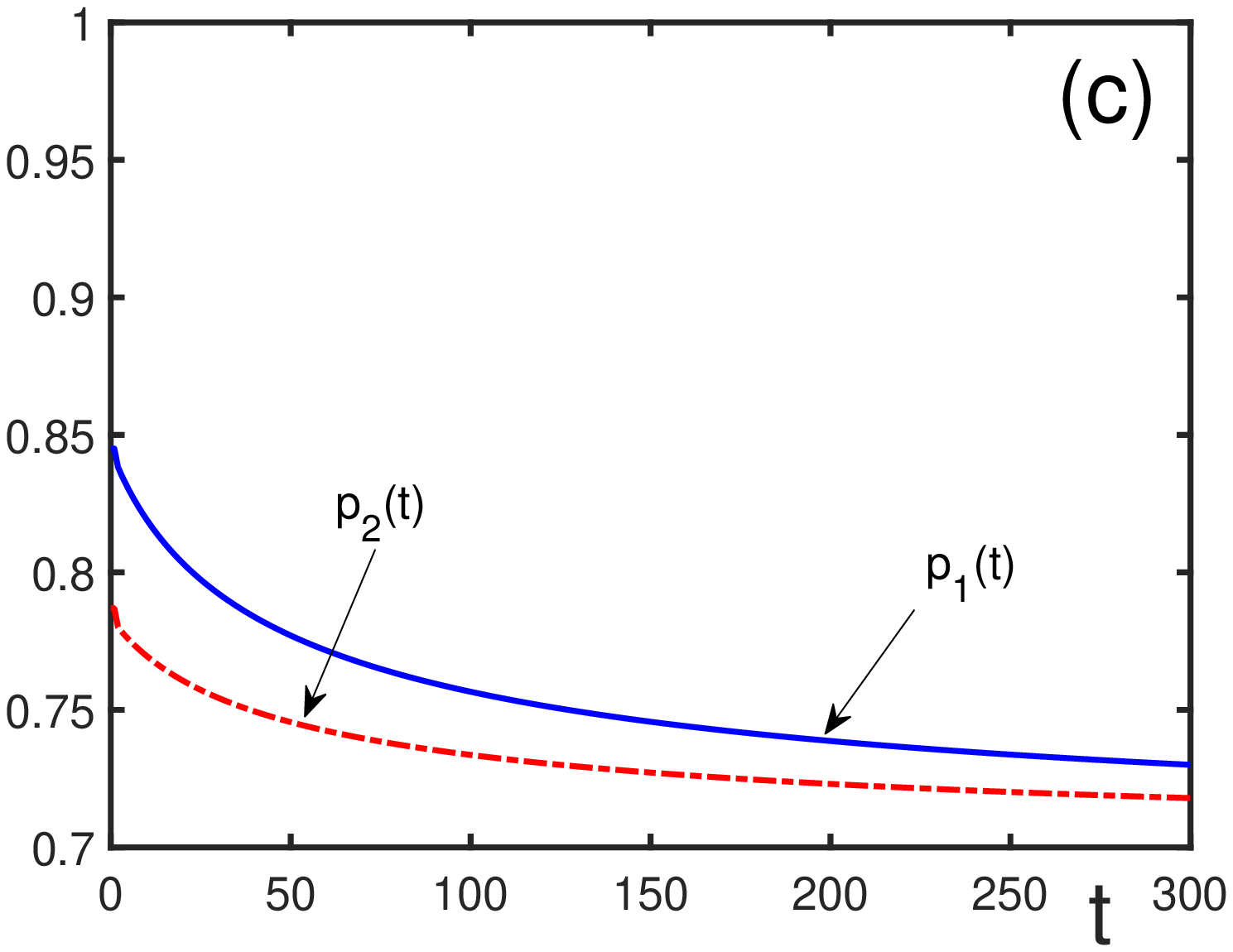}   } }
\caption{Probabilities $p_1(t)$ (solid line) and $p_2(t)$ (dash-dotted line) 
for the initial conditions $f_1= 0.4$, $f_2=0.1$, $q_1=0.59$, and $q_2=0.6$. 
(a) $\ep_1=\ep_2=0$. The fixed points are $p_1^*=0.4$ and $p_2^*=0.636$; 
(b) $\ep_1=0.5$ and $\ep_2=0$. The consensual fixed point is $p_1^*=p_2^*=0.7$; 
(c) $\ep_1=0.5$ and $\ep_2=0.3$. The consensual fixed point is $p_1^*=p_2^*=0.7$. 
}
\label{fig:Fig.1}
\end{figure}

\vskip 2mm
{\bf 2}. Without herding, the fraction $p_1(t)$ of agents with long-term memory tends 
to a fixed point without sharp variations, while the fraction $p_2(t)$ of agents with 
short-term memory experiences at the beginning essential oscillations that attenuate 
and finally the function tends to a fixed point, as is demonstrated in Fig. 2. 

%Figure 2
\begin{figure}[ht]
\centerline{
\hbox{ \includegraphics[width=7.5cm]{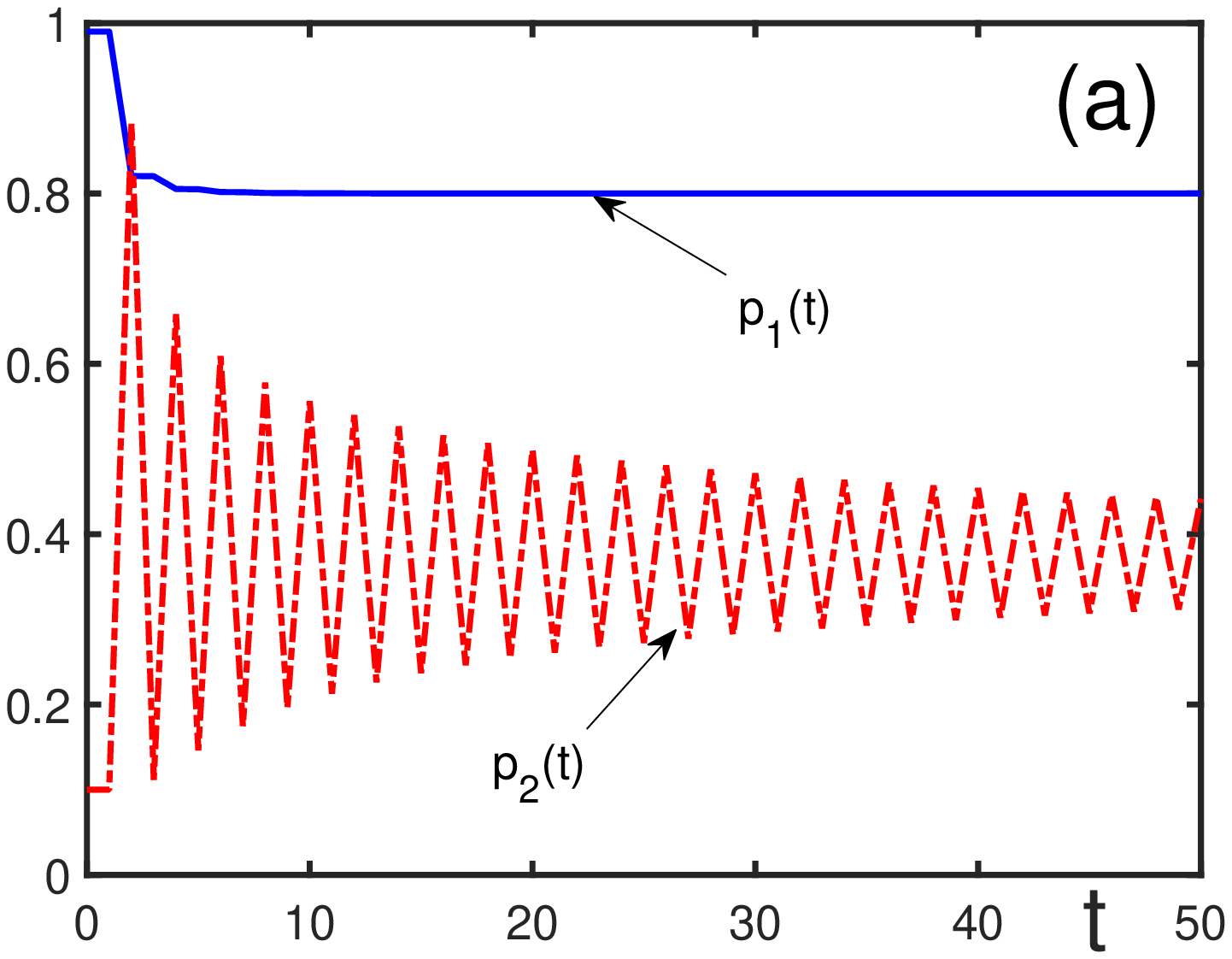} \hspace{1cm}
\includegraphics[width=7.5cm]{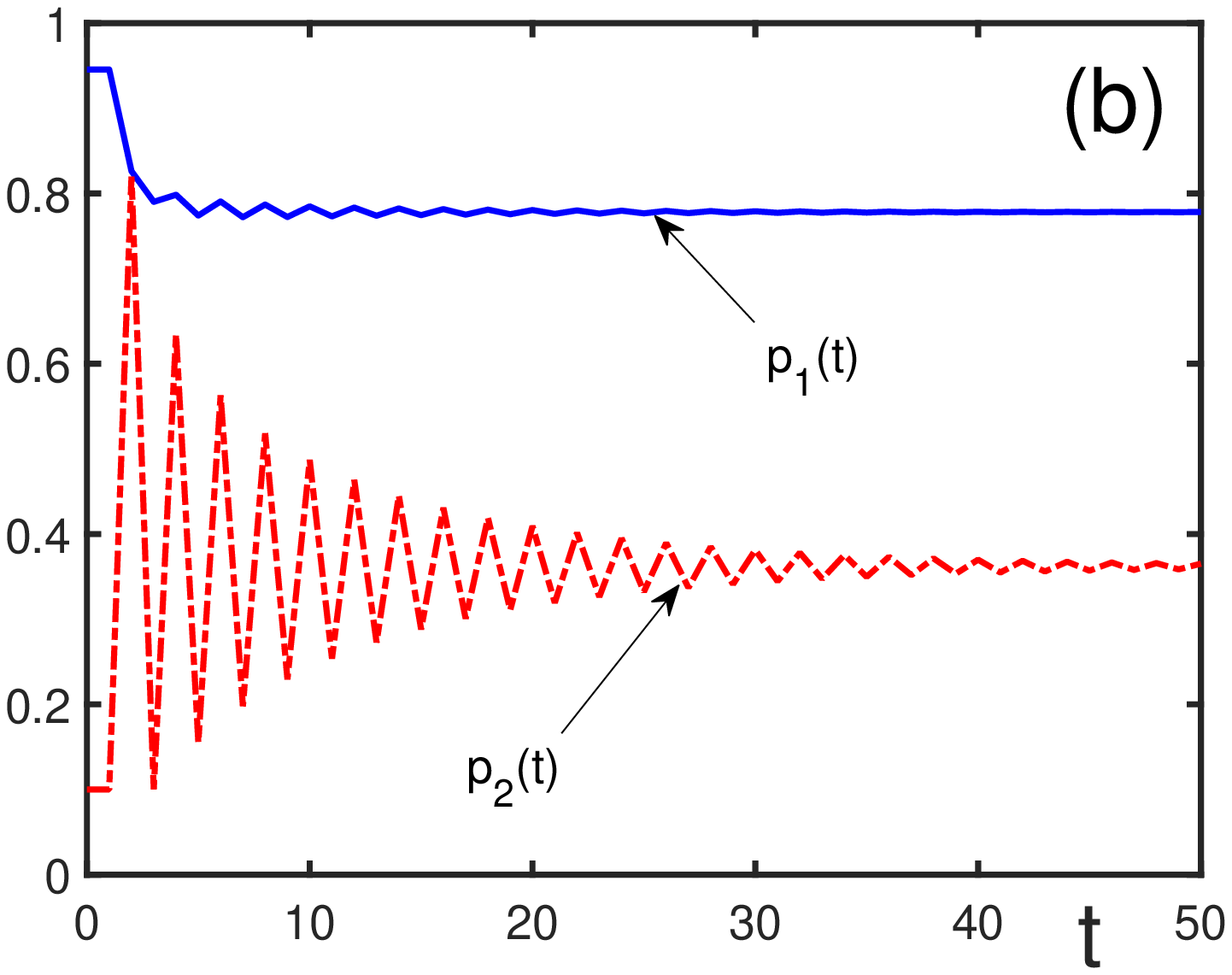}  } }
\vspace{12pt}
\centerline{
\hbox{ \includegraphics[width=7.5cm]{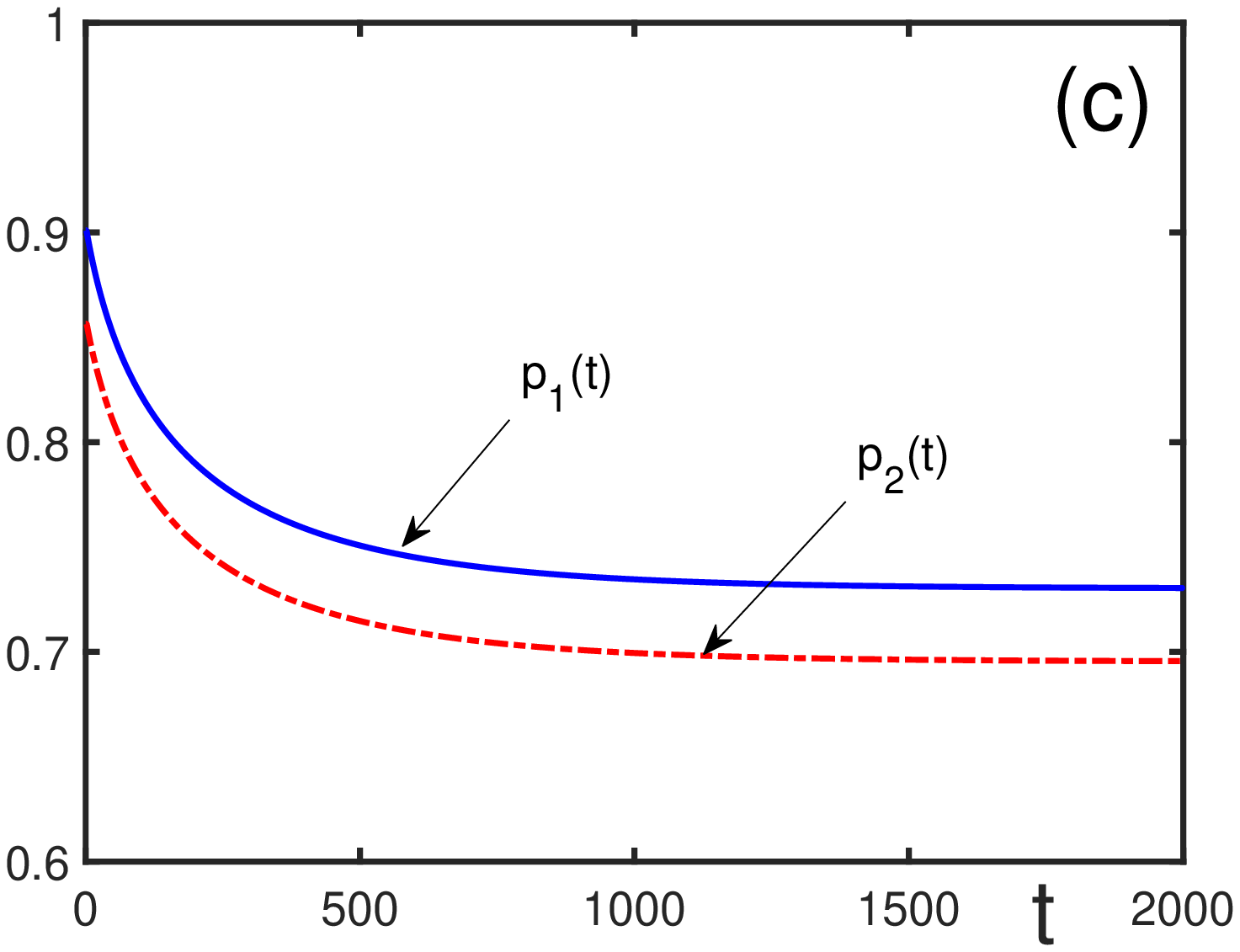} } }
\caption{Probabilities $p_1(t)$ (solid line) and $p_2(t)$ (dash-dotted 
line) for the initial conditions $f_1=0.8$, $f_2=0.9$, $q_1=0.19$, and 
$q_2=-0.8$.
(a) $\ep_1=\ep_2= 0$. The fixed points are $p_1^*= 0.8$ and $p_2^*=0.377$; 
(b) $\ep_1=0.05$ and $\ep_2=0$. The fixed points are $p_1^*=0.778$ and 
$p_2^*=0.362$; 
(c) $\ep_1=0.1$ and $\ep_2=0.85$. The fixed points are $p_1^*=0.730$ and 
$p_2^*=0.695$.
}
\label{fig:Fig.2}
\end{figure}

\vskip 2mm
{\bf 3}. When there is no herding, $p_1(t)$ tends to a fixed point, while $p_2(t)$
permanently oscillates. Switching on the herding parameters results in the permanent
oscillation of both functions $p_1(t)$, as well as $p_2(t)$. Increasing the herding
parameters further, after initial oscillations, forces both functions to tend to their
fixed points. This is illustrated in Fig. 3.

%Figure 3
\begin{figure}[ht]
\centerline{
\hbox{ \includegraphics[width=7.5cm]{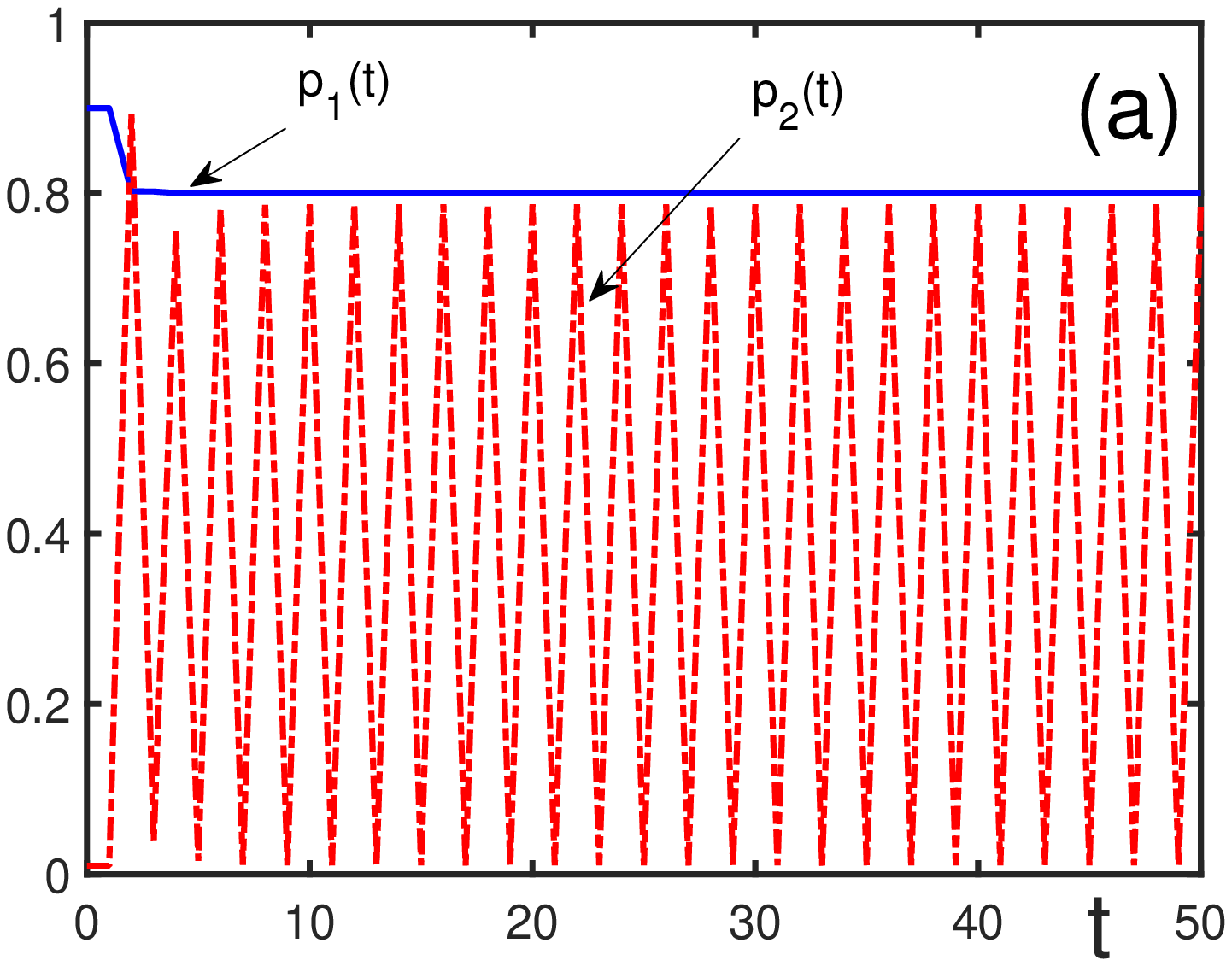} \hspace{1cm}
\includegraphics[width=7.5cm]{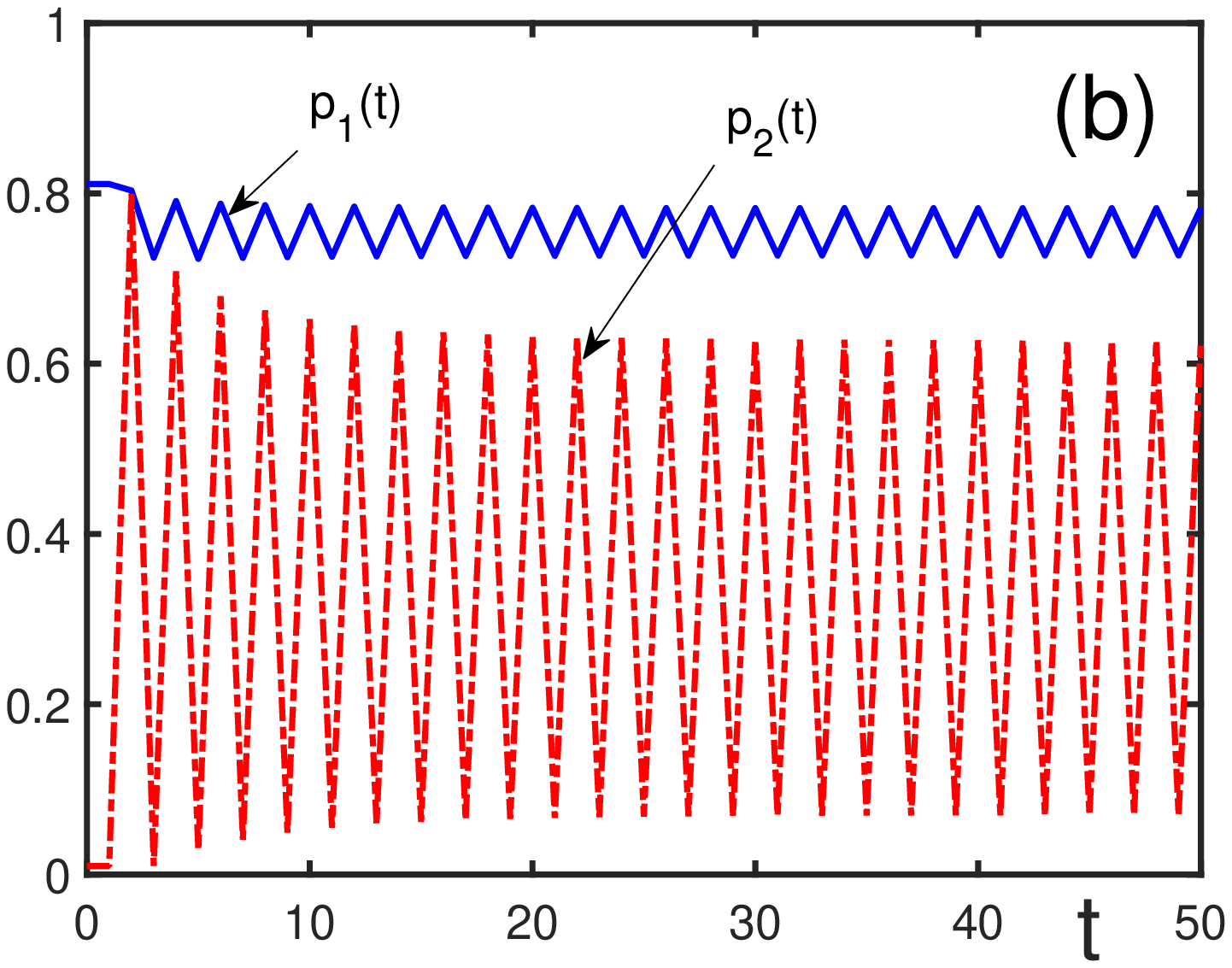}  } }
\vspace{12pt}
\centerline{
\hbox{ \includegraphics[width=7.5cm]{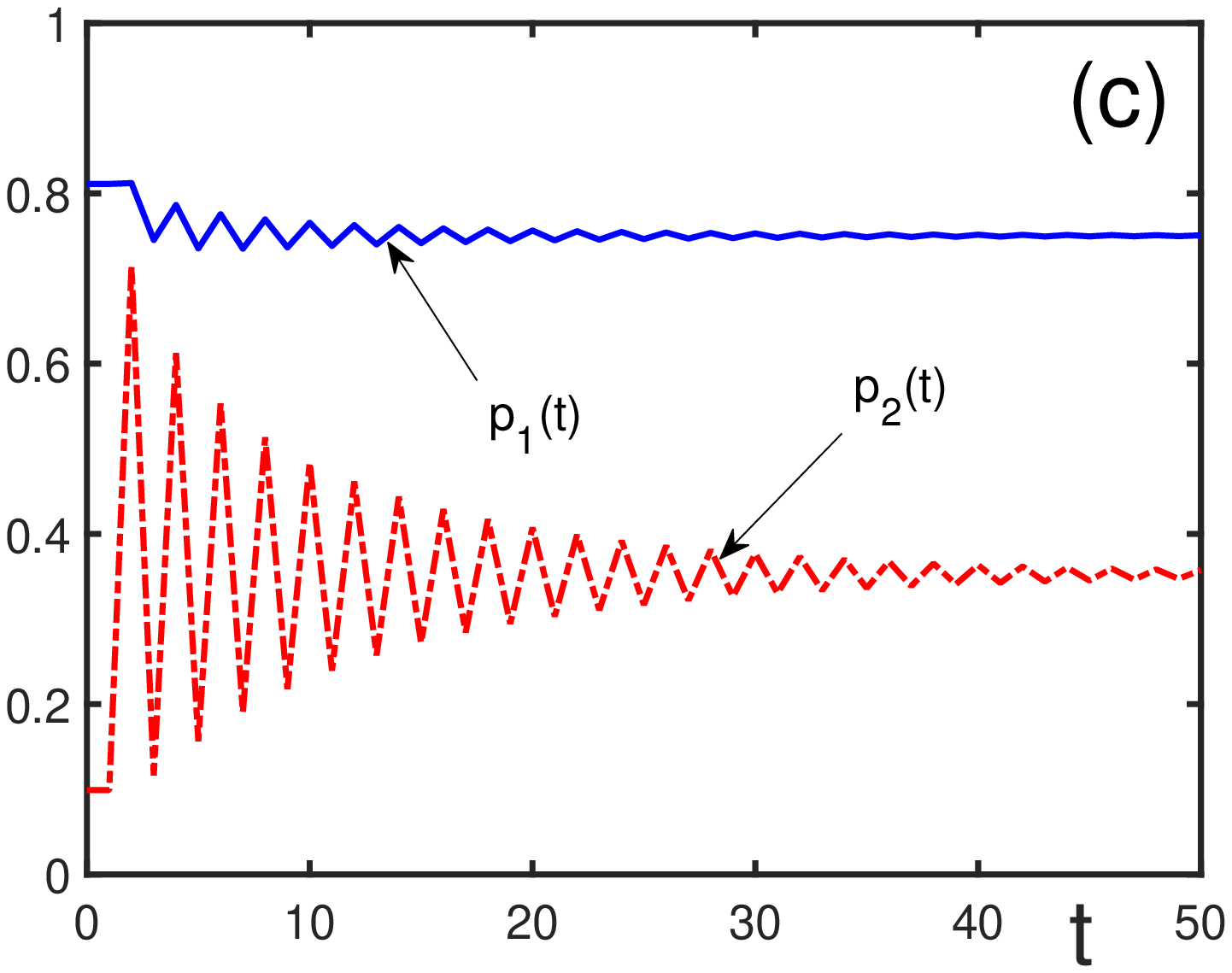} \hspace{1cm}
\includegraphics[width=7.5cm]{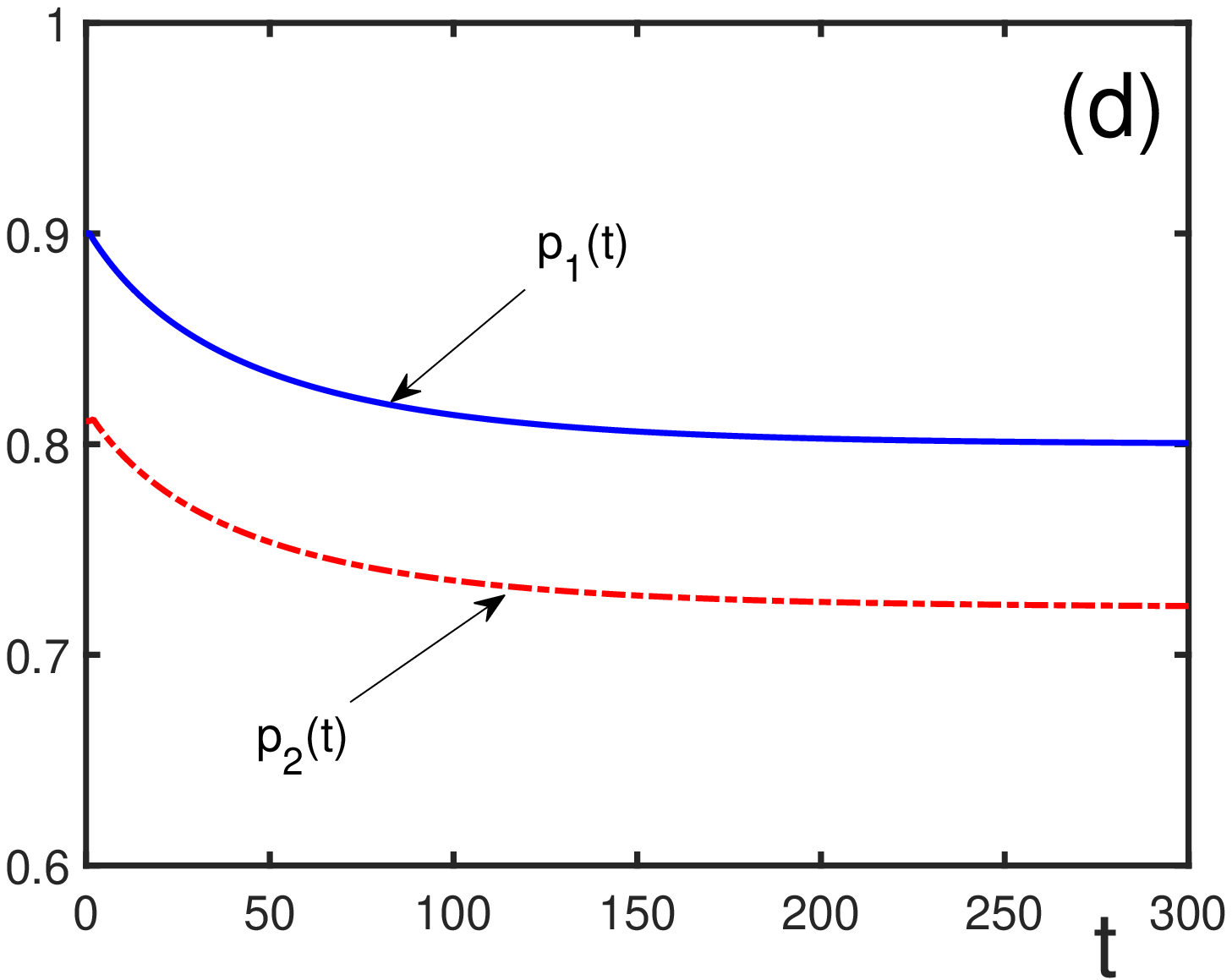}  } }
\caption{Probabilities $p_1(t)$ (solid line) and $p_2(t)$ (dash-dotted line) 
for the initial conditions $f_1=0.8$, $f_2=1$, $q_1=0.1$, and $q_2=-0.99$.
(a) $\ep_1=\ep_2=0$. The probability for the agents with long-term memory tends 
to the fixed point $p_1^*=0.8$, while the function $p_2(t)$ oscillates for all 
times; 
(b) $\ep_1=0.1$ and $\ep_2=0$. Both functions $p_1(t)$ and $p_2(t)$ permanently 
oscillate; 
(c) $\ep_1=\ep_2=0.1$. The functions, after experiencing at the beginning 
oscillations, tend to the fixed points $p_1^*=0.75$ and $p_2^*=0.353$; 
(d) $\ep_1=0$ and $\ep_2=0.9$. Both probabilities tend to the fixed points 
$p_1^*=0.8$ and $p_2^*=0.723$.
}
\label{fig:Fig.3}
\end{figure}  

\vskip 2mm
{\bf 4}. When the herding effect is absent, $p_1(t)$ monotonically tends to a fixed 
point, but $p_2(t)$ exhibits an unusual behaviour. At the beginning, for sufficiently 
long time, $p_2(t)$ behaves smoothly. Then suddenly, it starts widely oscillating and 
continues oscillating for ever. Slightly increasing the herding parameters shifts the 
beginning of oscillations to larger times, makes the oscillation amplitude smaller, and 
makes the function $p_1(t)$ to also experience everlasting oscillations. The following 
increase of the herding parameters eliminates oscillations and leads to smooth tendency 
of both probabilities to different fixed points. The further growth of the herding 
parameters forces both functions to converge to a consensual fixed point. See Fig. 4.  

%Figure 4
\begin{figure}[ht]
\centerline{
\hbox{ \includegraphics[width=7.5cm]{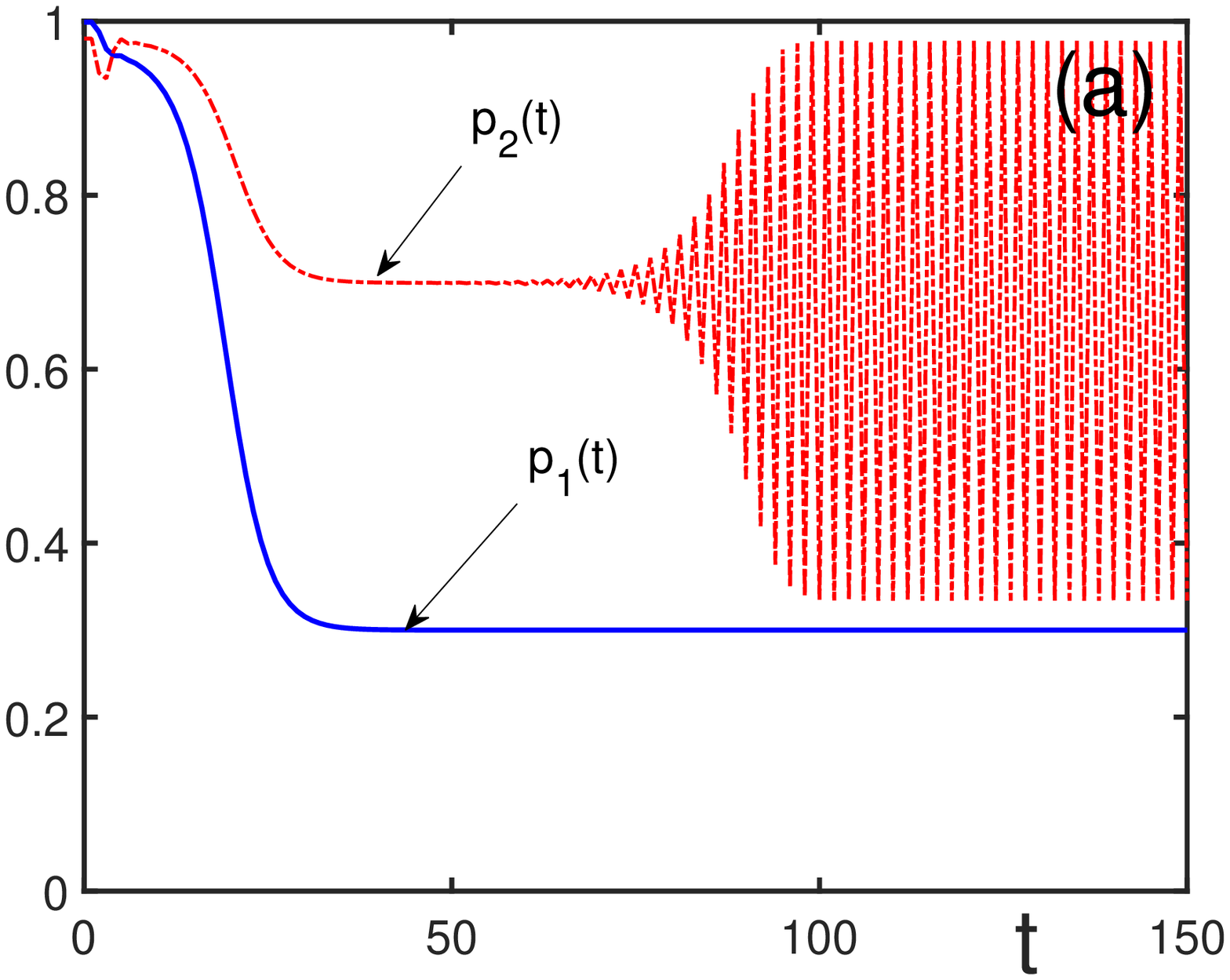} \hspace{1cm}
\includegraphics[width=7.5cm]{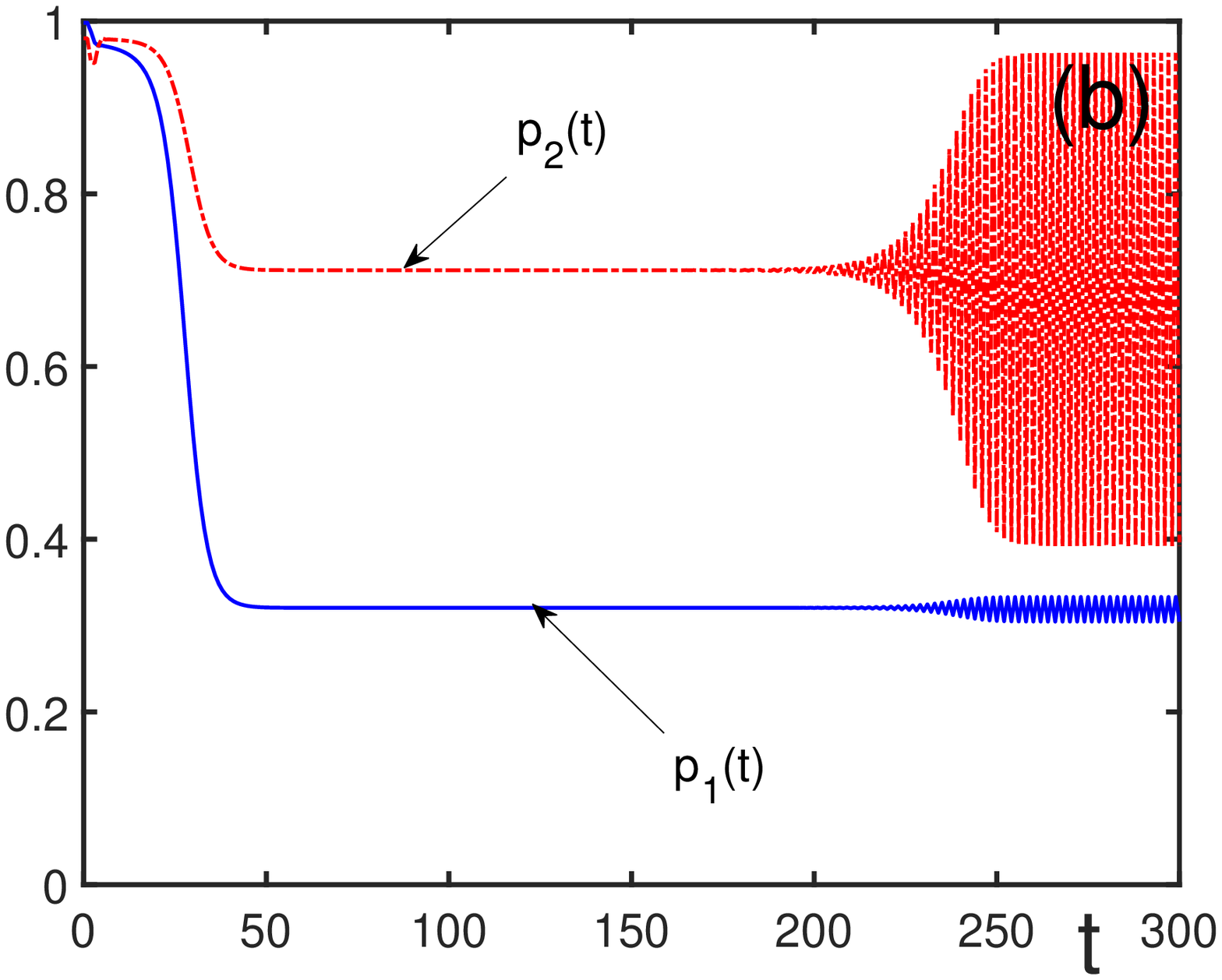}  } }
\vspace{12pt}
\centerline{
\hbox{ \includegraphics[width=7.5cm]{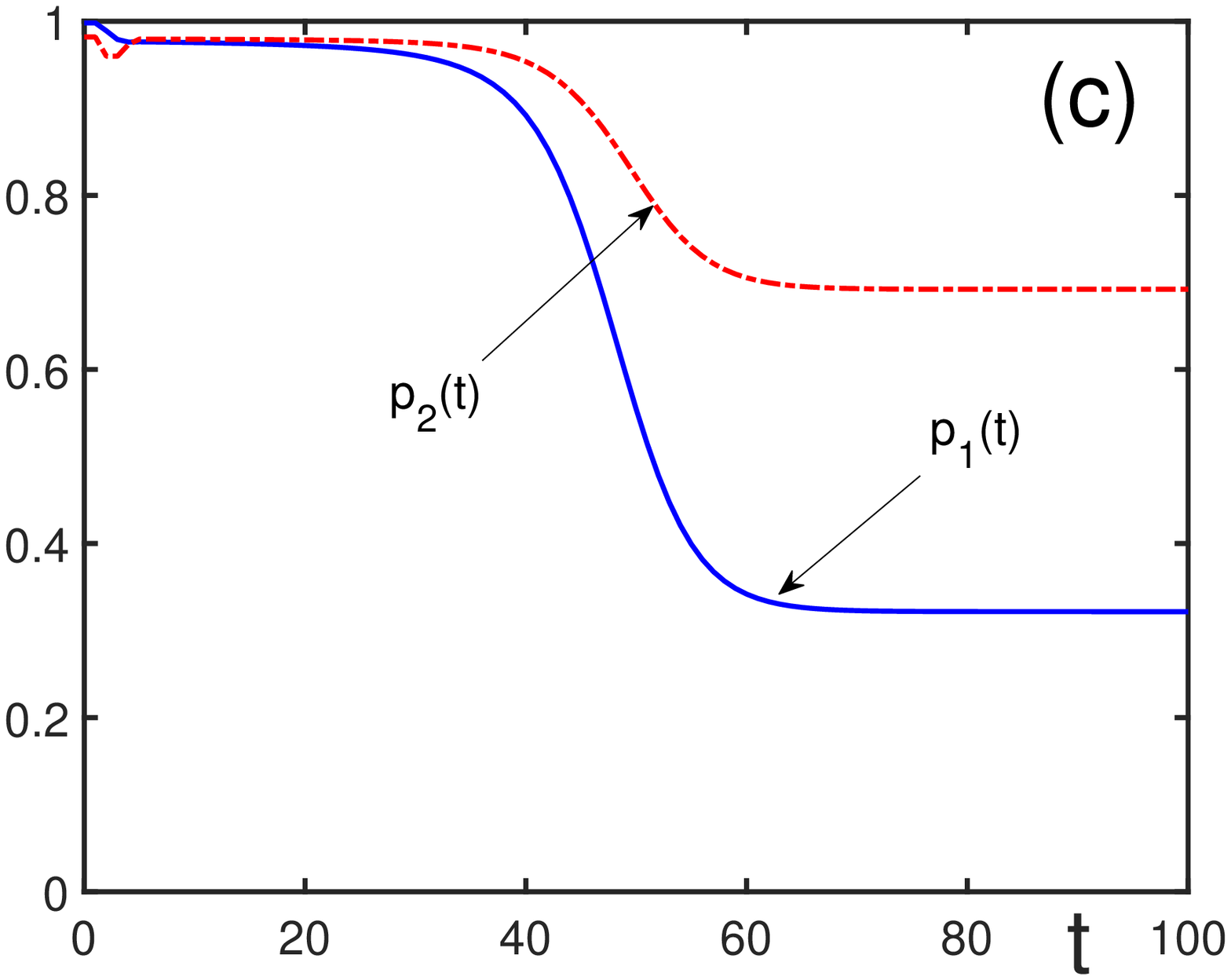} \hspace{1cm}
\includegraphics[width=7.5cm]{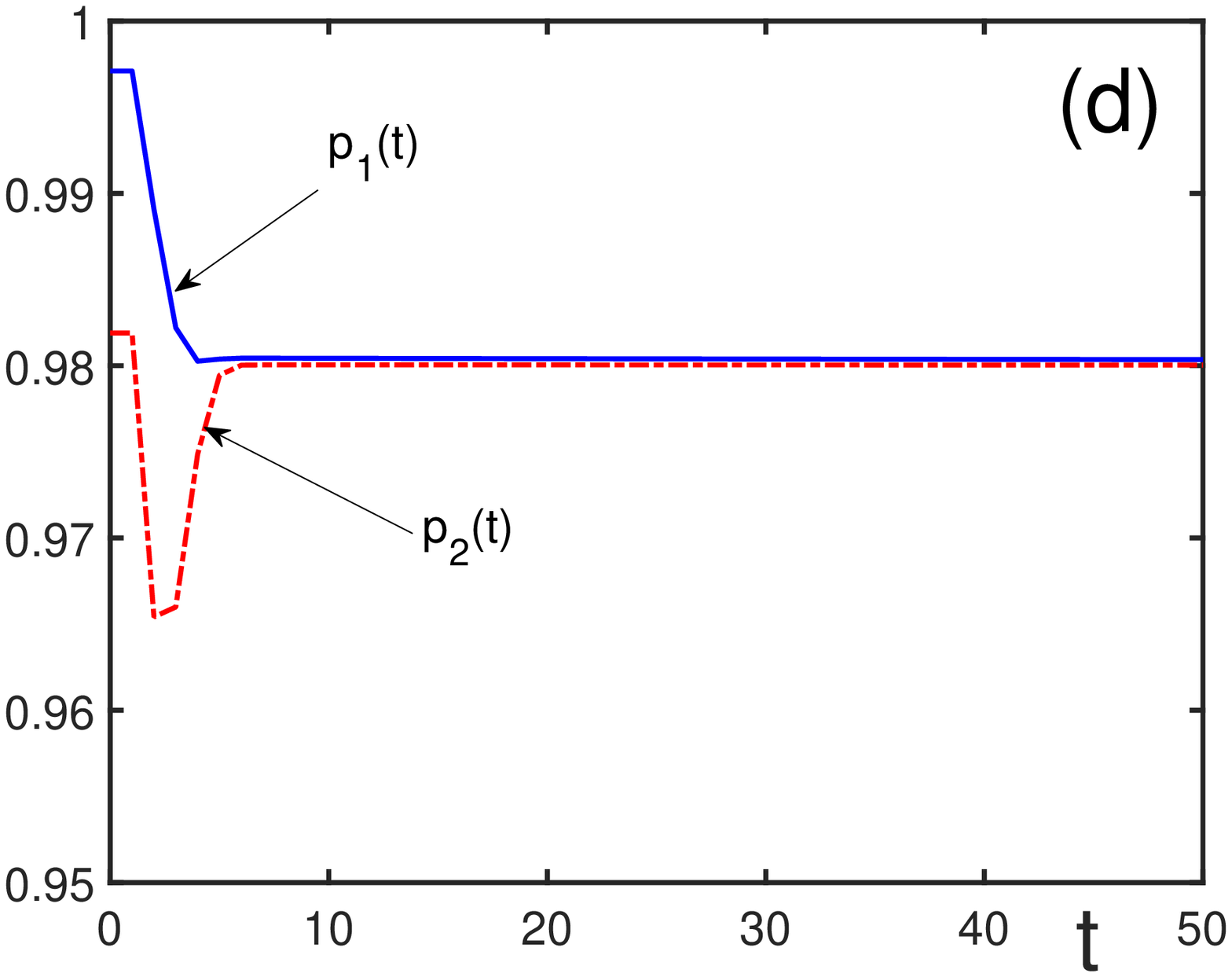} } }
\vspace{12pt}
\centerline{
\hbox{ \includegraphics[width=7.5cm]{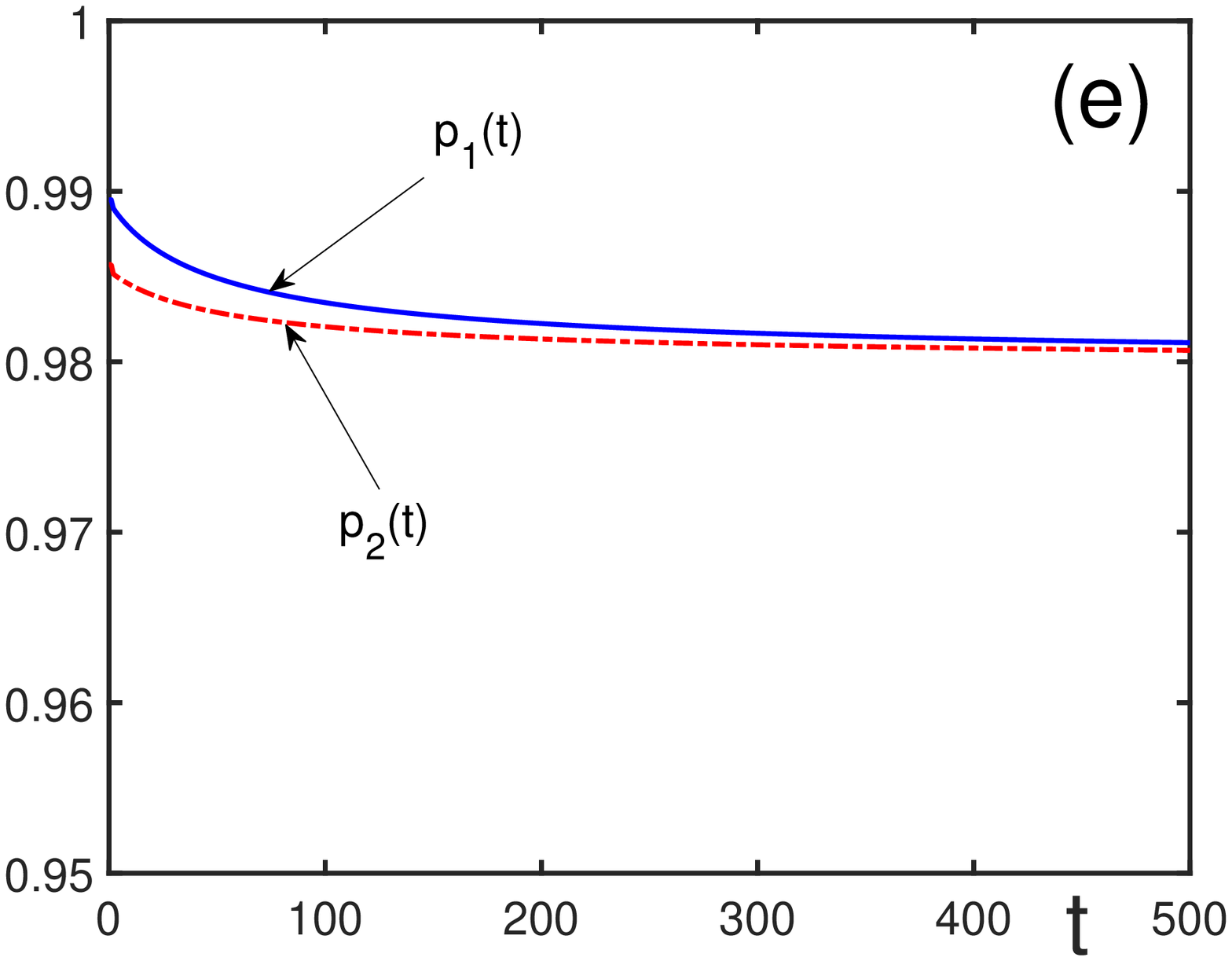} } }
\caption{Probabilities $p_1(t)$ (solid line) and $p_2(t)$ (dash-dotted line) 
for the initial conditions $f_1=0.3$, $f_2=0$, $q_1=0.699$, and $q_2=0.98$.
(a) $\ep_1=\ep_2=0$. The function $p_1(t)$ tends to the fixed point $p_1^*=0.3$, 
while $p_2(t)$, after sufficiently long time of smooth behaviour, exhibits 
everlasting oscillations; 
(b) $\ep_1=0.05$ and $\ep_2=0$. Both functions $p_1(t)$ and $p_2(t)$ always 
oscillate; 
(c) $\ep_1=0.05$ and $\ep_2=0.1$. Both  functions tend to different fixed points 
$p_1^*=0.322$ and $p_2^*=0.692$; 
(d) $\ep_1=\ep_2=0.1$. Both probabilities tend to the consensual fixed point 
$p_1^*=p_2^*=0.98$; 
(e) $\ep_1= 0.5$ and $\ep_2=0.3$; Both probabilities $p_1(t)$ and $p_2(t)$ 
monotonically tend to the consensual fixed point $p_1^*=p_2^*=0.98$. This is 
an example showing the spontaneous appearance of self-excited oscillations that 
can be suppressed by the herding effect. 
}
\label{fig:Fig.4}
\end{figure}

\vskip 2mm
Figures 5 to 10 illustrate the case of large herding parameters, such that 
$\ep_1+\ep_2>1$. We start with the values $\ep_1=\ep_2=1$, so that $\ep_1+\ep_2=2$, 
when the dynamics is described by Eqs. (\ref{36}) and then consider diminishing 
parameters approaching the line $\ep_1+\ep_2=1$, where Eq. (\ref{35}) is valid. The 
line $\ep_1+\ep_2=1$ corresponds to the case of the strongest herding behavior.

\vskip 2mm
{\bf 5}. For $\ep_1+\ep_2=2$, the probabilities tend to different fixed points. However, 
diminishing the herding parameters to the line $\ep_1+\ep_2=1$ results in the convergence 
of both functions $p_j(t)$ to the common fixed point, as is seen from Fig. 5.  

%Figure 5
\begin{figure}[ht]
\centerline{
\hbox{ \includegraphics[width=7.5cm]{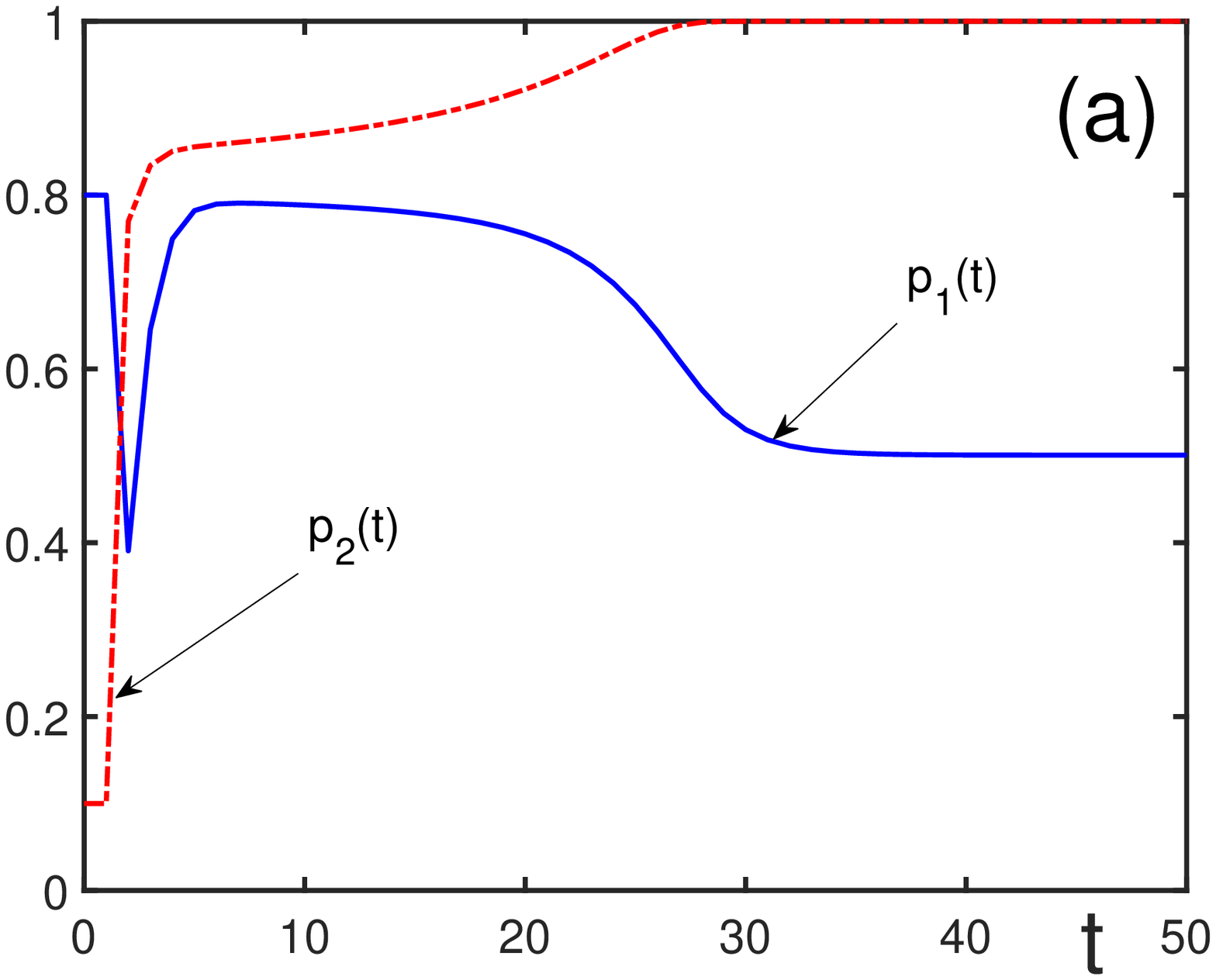} \hspace{1cm}
\includegraphics[width=7.5cm]{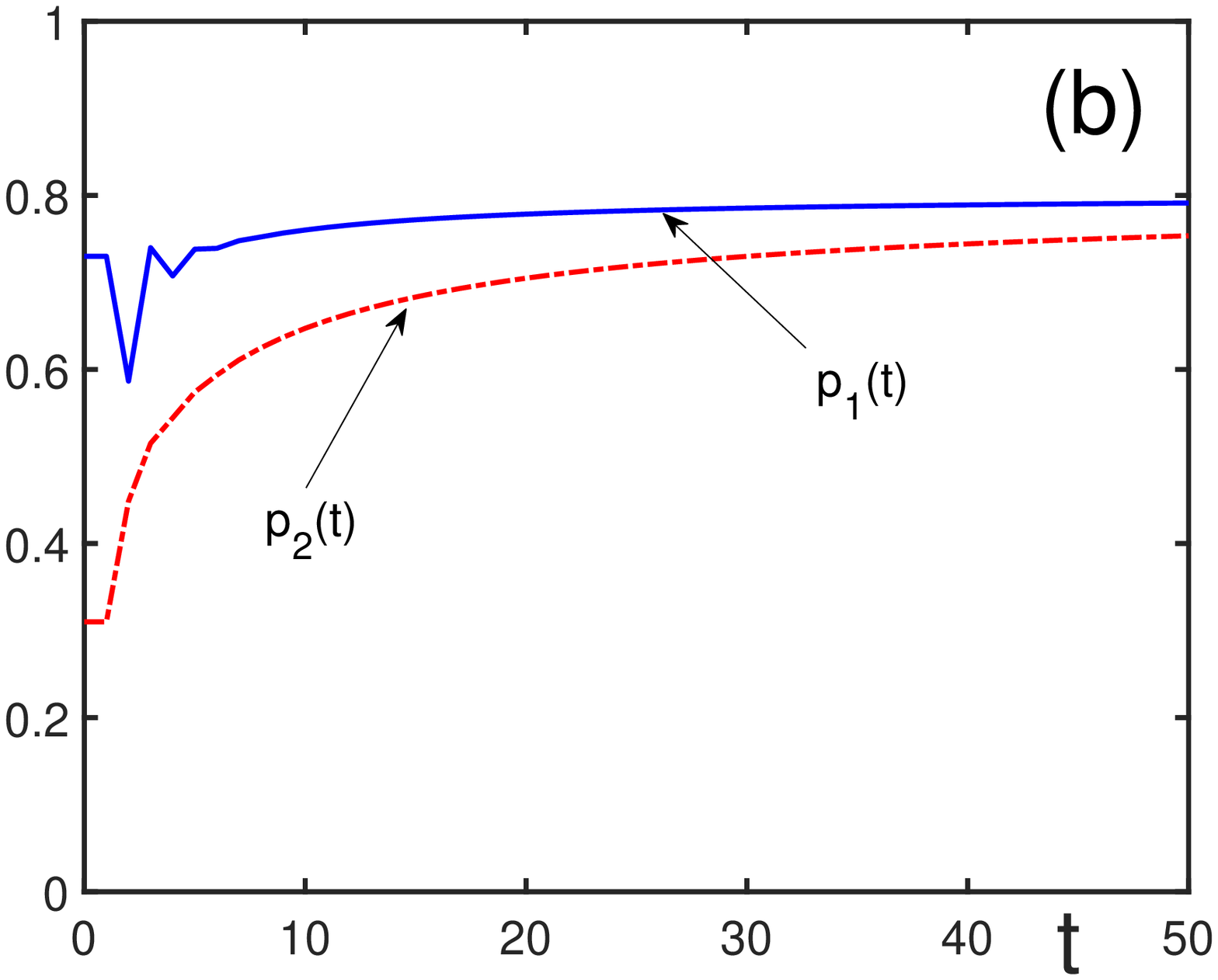}  } }
\vspace{12pt}
\centerline{
\hbox{ \includegraphics[width=7.5cm]{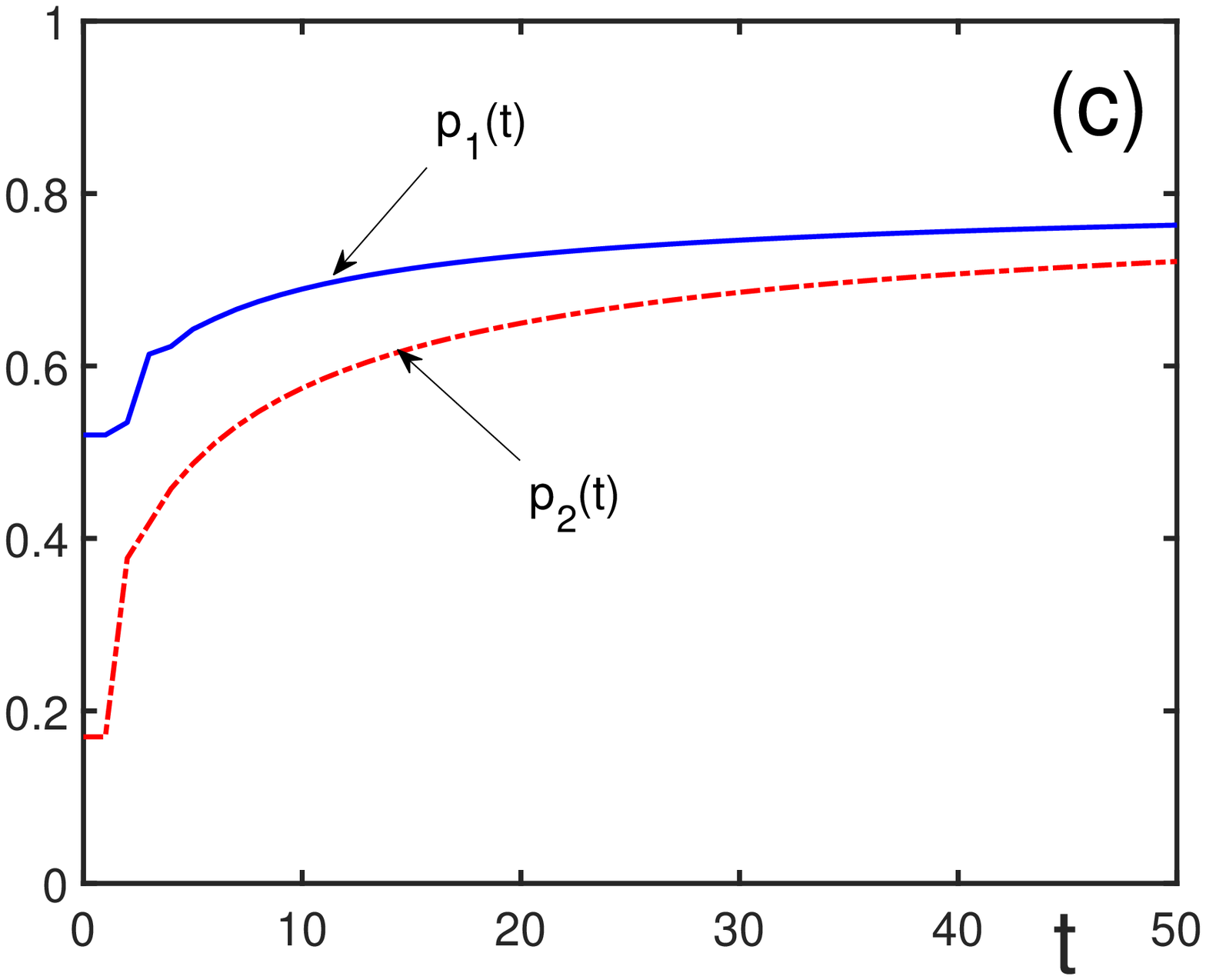} \hspace{1cm}
\includegraphics[width=7.5cm]{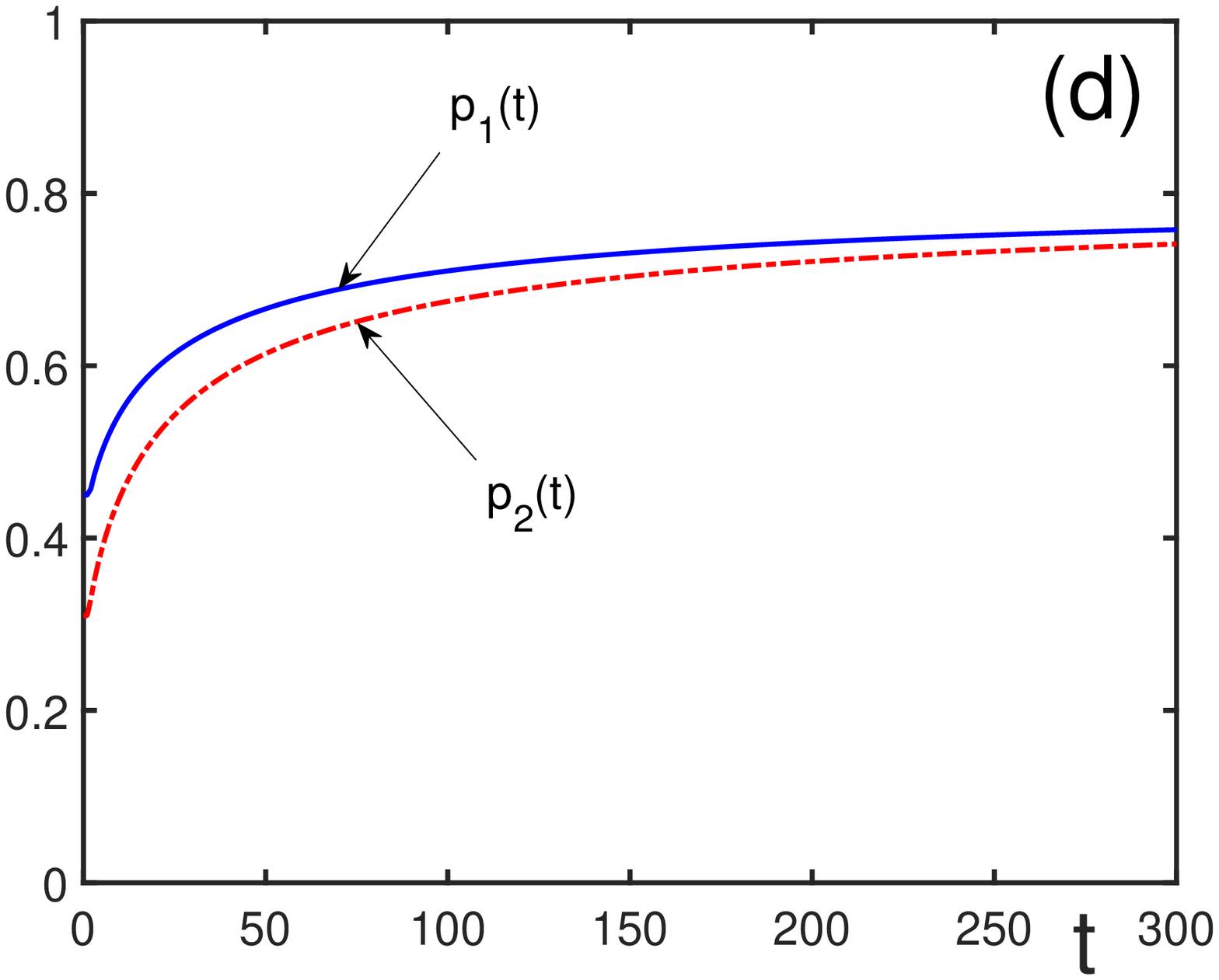}  } }
\caption{Probabilities $p_1(t)$ (solid line) and $p_2(t)$ (dash-dotted line) for 
the initial conditions $f_1=1$, $f_2=0.2$, $q_1=-0.9$, and $q_2=0.6$.
(a) $\ep_1=\ep_2=1$. The fixed points are $p_1^*=0.5$ and $p_2^*=1$; 
(b) $\ep_1=0.9$ and $\ep_2=0.7$. The consensual fixed point is $p_1^*=p_2^*=0.8$; 
(c) $\ep_1=0.6$ and $\ep_2=0.9$. The consensual fixed point is $p_1^*=p_2^*=0.8$; 
(d) $\ep_1=0.5$ and $\ep_2=0.7$. Both functions tend to $p_1^*=p_2^*=0.8$. 
}
\label{fig:Fig.5}
\end{figure}

\vskip 2mm
{\bf 6}. When $\ep_1=\ep_2=1$, the probability $p_1(t)$ for the agents with long-term 
memory oscillates from the beginning, while $p_2(t)$ tends monotonically to a fixed point. 
This is contrary to Fig. 3, which is in agreement with Eqs. (\ref{36}). Diminishing the 
herding parameters, first, makes both functions permanently oscillating, but then 
suppresses the oscillations, so that close to the line $\ep_1+\ep_2=1$ both probabilities 
tend to fixed points, as is shown in Fig. 6.

%Figure 6
\begin{figure}[ht]
\centerline{
\hbox{ \includegraphics[width=7.5cm]{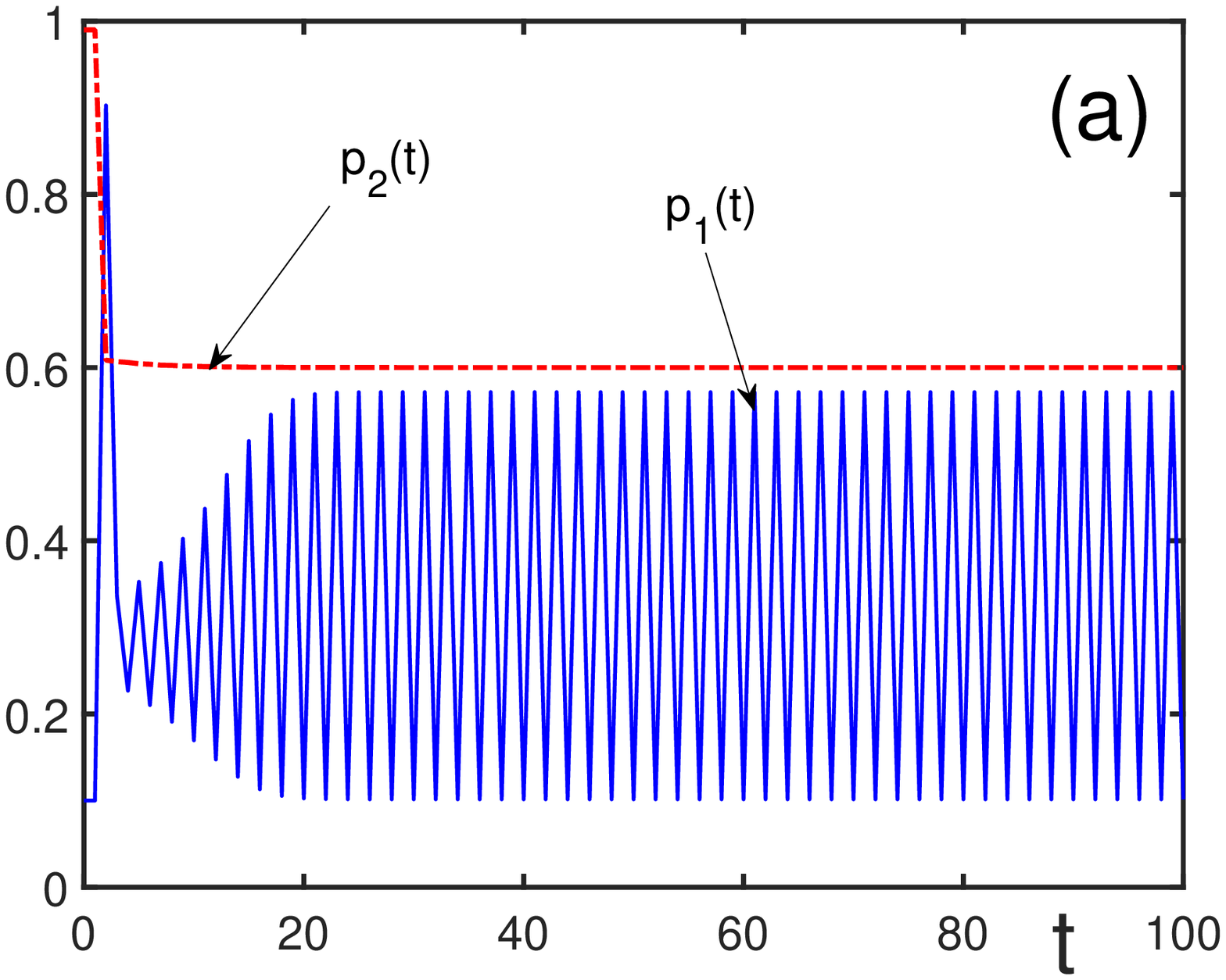} \hspace{1cm}
\includegraphics[width=7.5cm]{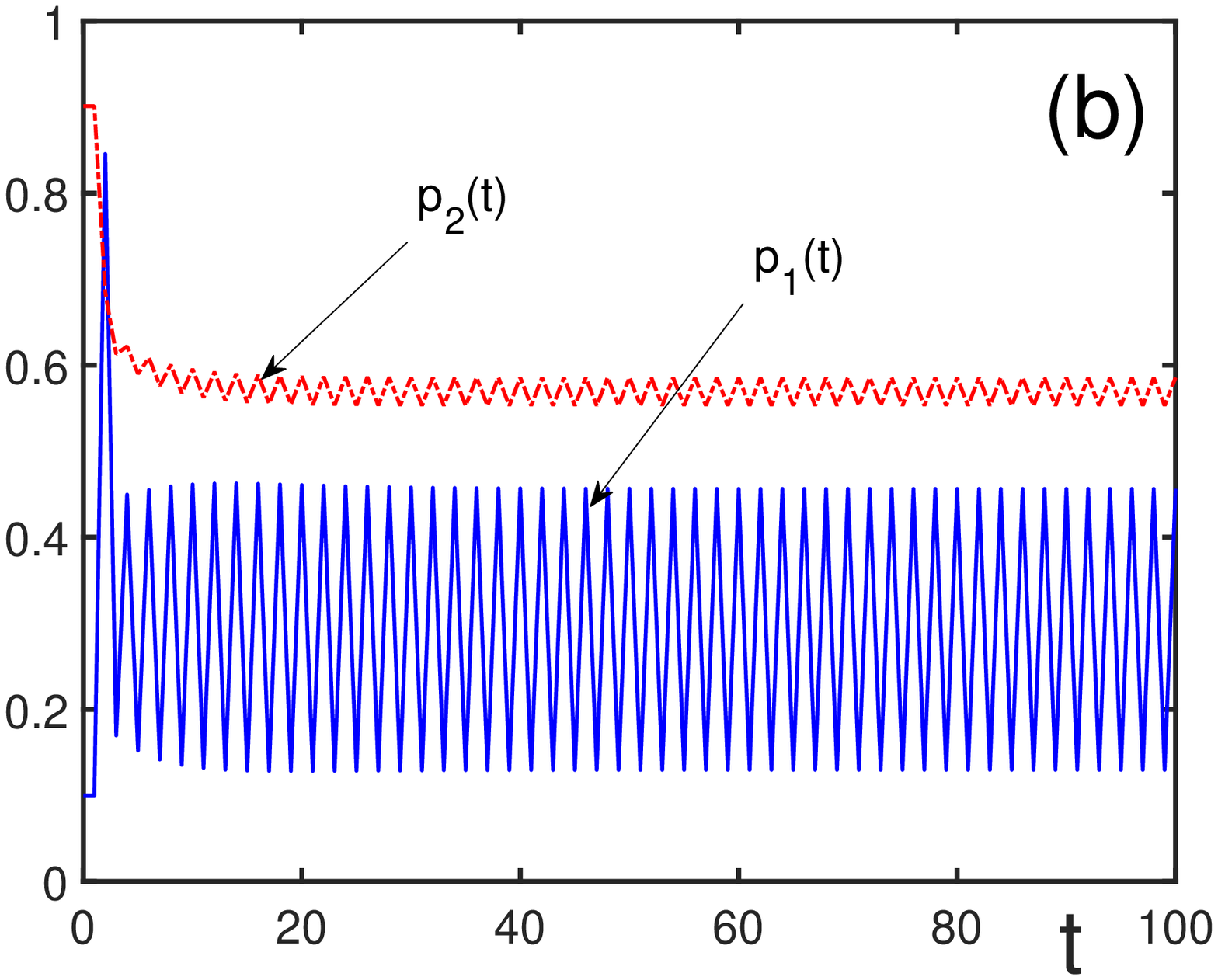}  } }
\vspace{12pt}
\centerline{
\hbox{ \includegraphics[width=7.5cm]{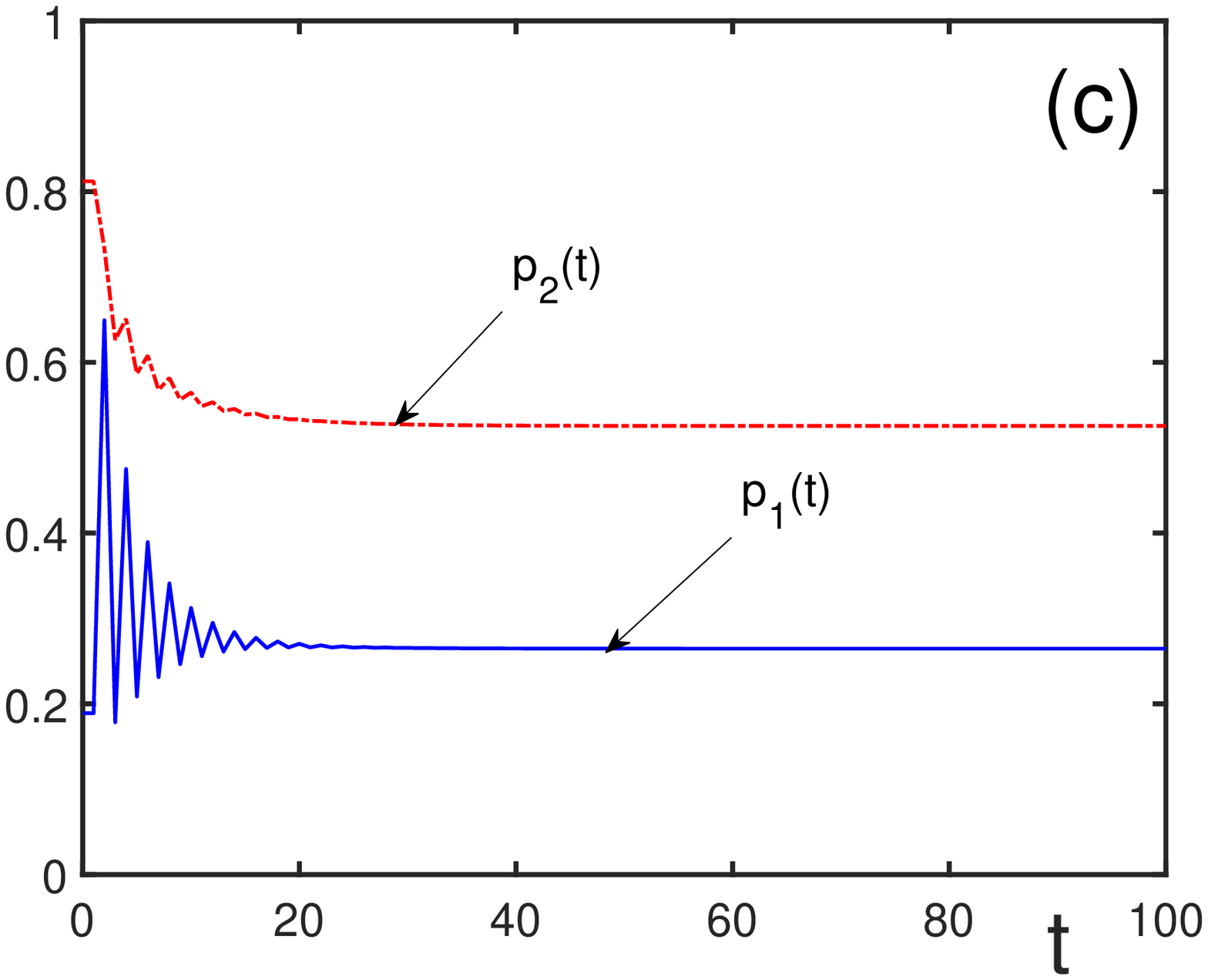} \hspace{1cm}
\includegraphics[width=7.5cm]{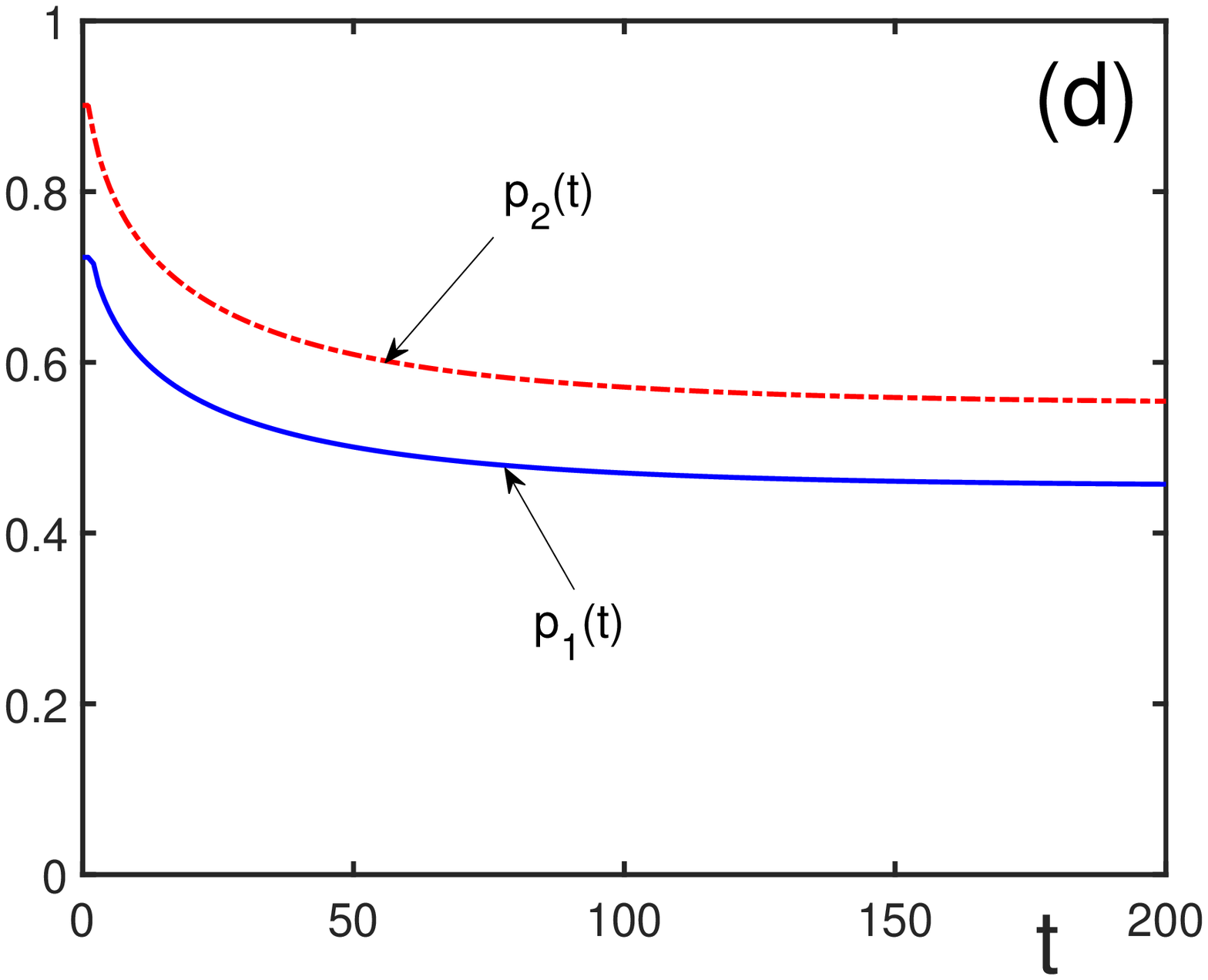}  } }
\caption{Probabilities $p_1(t)$ (solid line) and $p_2(t)$ (dash-dotted line) 
for the initial conditions $f_1=0.6$, $f_2=1$, $q_1=0.39$, and $q_2=-0.9$.
(a) $\ep_1=\ep_2=1$. The function $p_1(t)$ oscillates with a constant amplitude 
and $p_2(t)$ tends monotonically to the fixed point $p_2^*=0.6$;
(b) $\ep_1=1$ and $\ep_2=0.9$. Both functions $p_1(t)$ and $p_2(t)$ permanently 
oscillate with constant amplitudes; 
(c) $\ep_1=0.9$ and $\ep_2=0.8$. The functions tend to the fixed points 
$p_1^*=0.265$ and $p_2^*=0.525$, respectively; 
(d) $\ep_1=0.3$ and $\ep_2=0.9$. The fixed points are $p_1^*=0.455$ and 
$p_2^*=0.552$. 
}
\label{fig:Fig.6}
\end{figure}
    
\vskip 2mm
{\bf 7}. For the case $\ep_1=\ep_2=1$, the probability $p_1(t)$ for the agents with 
long-term memory permanently oscillates, at the beginning a little, but then the 
oscillation amplitude increases, while $p_2(t)$ monotonically tends to a fixed point.
Diminishing the herding parameters leads to the oscillation of both probabilities and 
shifts the start of oscillations to larger times. Then, reducing the herding parameters
to the line $\ep_1+\ep_2=1$ suppresses the oscillations and forces both functions to 
converge, first to different fixed points and then to the same consensual fixed point, 
which is illustrated in Fig. 7.

%Figure 7
\begin{figure}[ht]
\centerline{
\hbox{ \includegraphics[width=7.5cm]{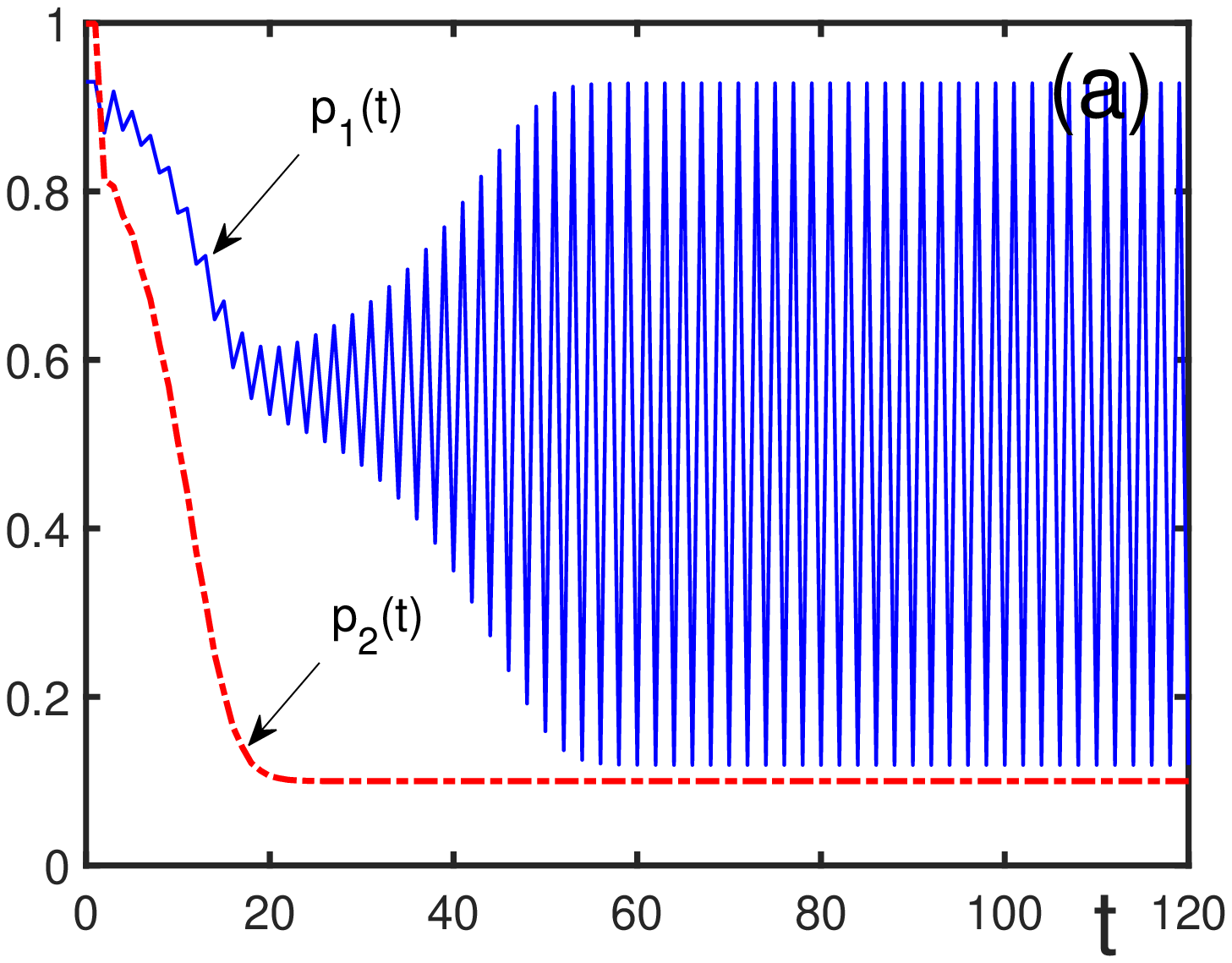} \hspace{1cm}
\includegraphics[width=7.5cm]{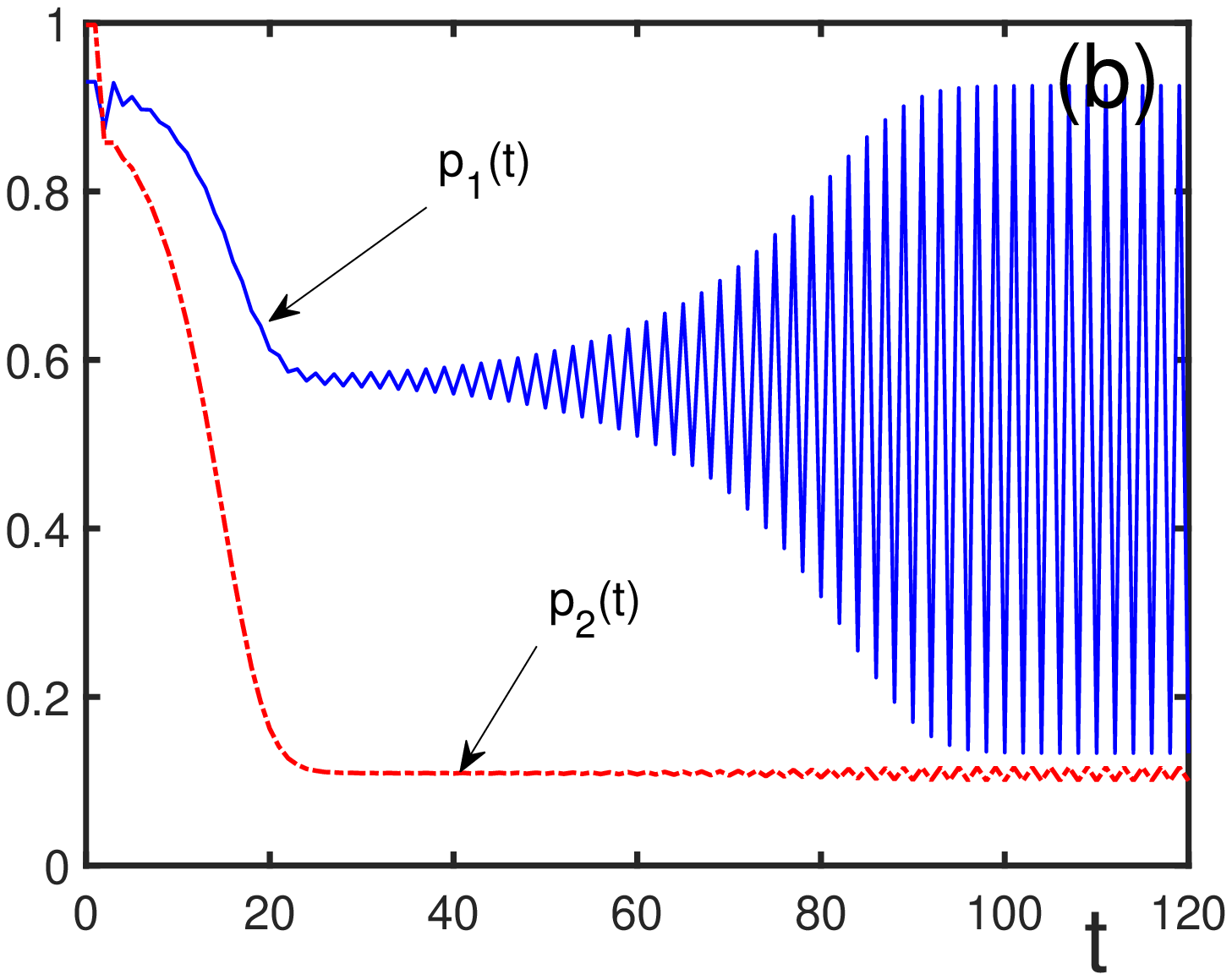}  } }
\vspace{12pt}
\centerline{
\hbox{ \includegraphics[width=7.5cm]{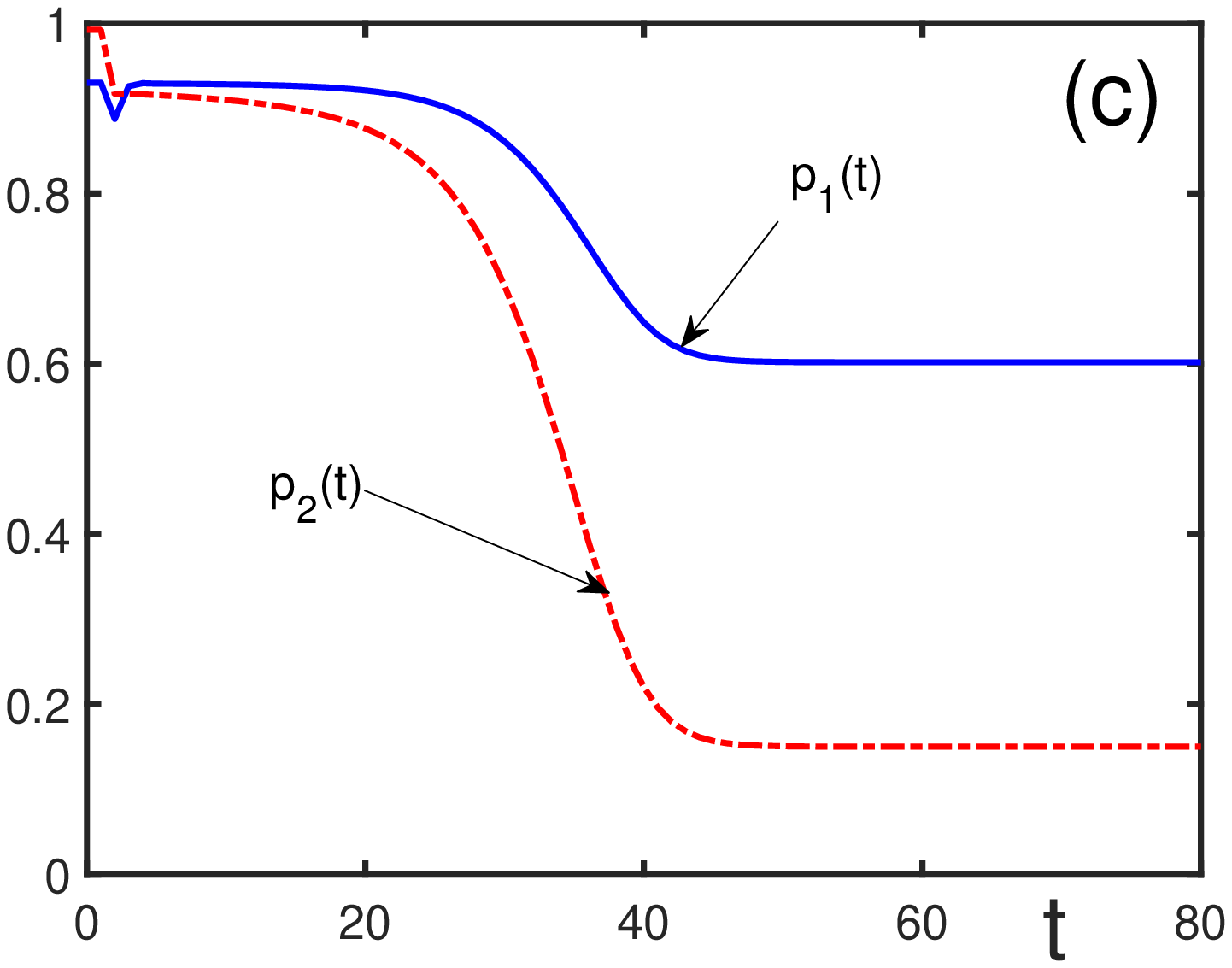} \hspace{1cm}
\includegraphics[width=7.5cm]{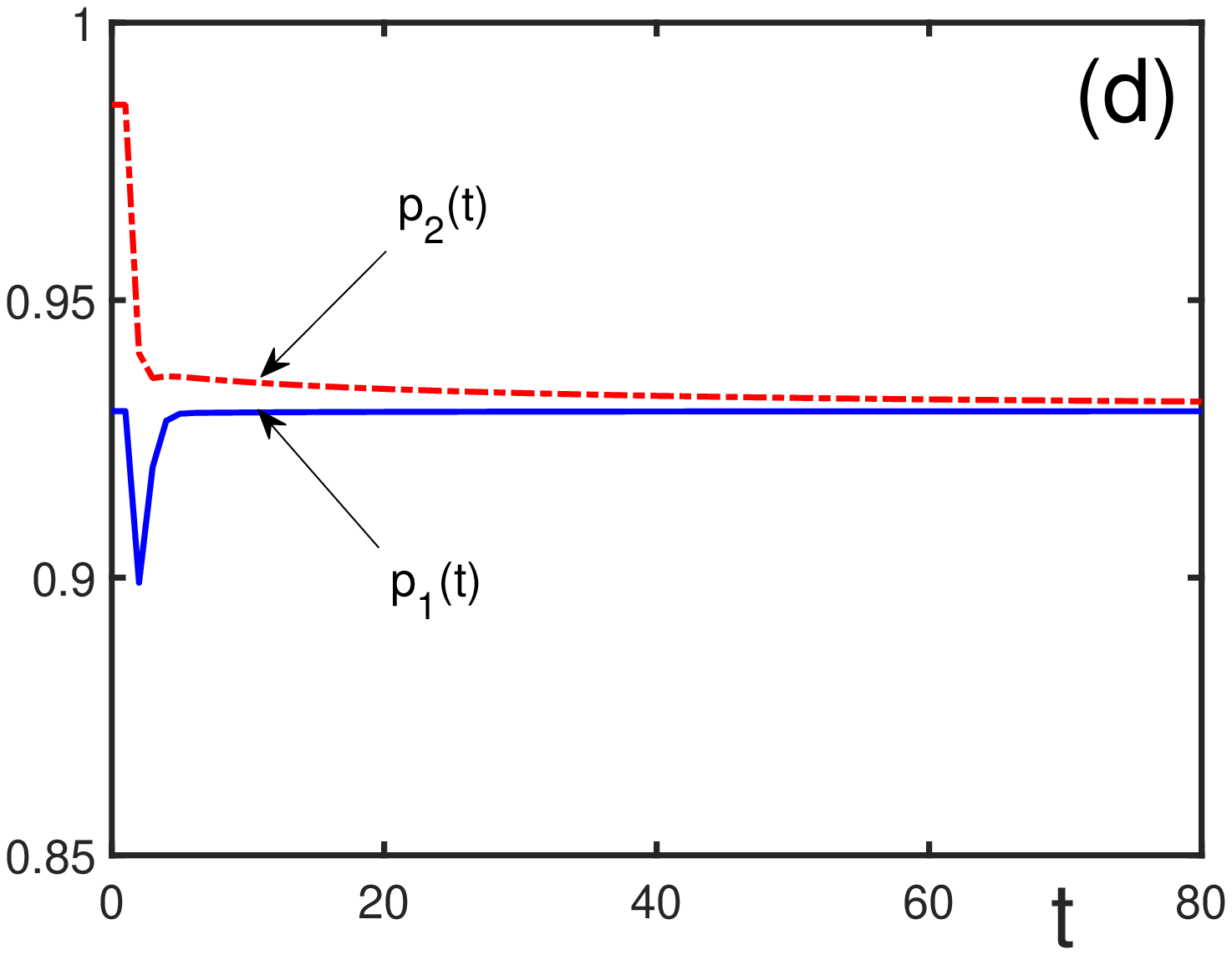} } }
\vspace{12pt}
\centerline{
\hbox{ \includegraphics[width=7.5cm]{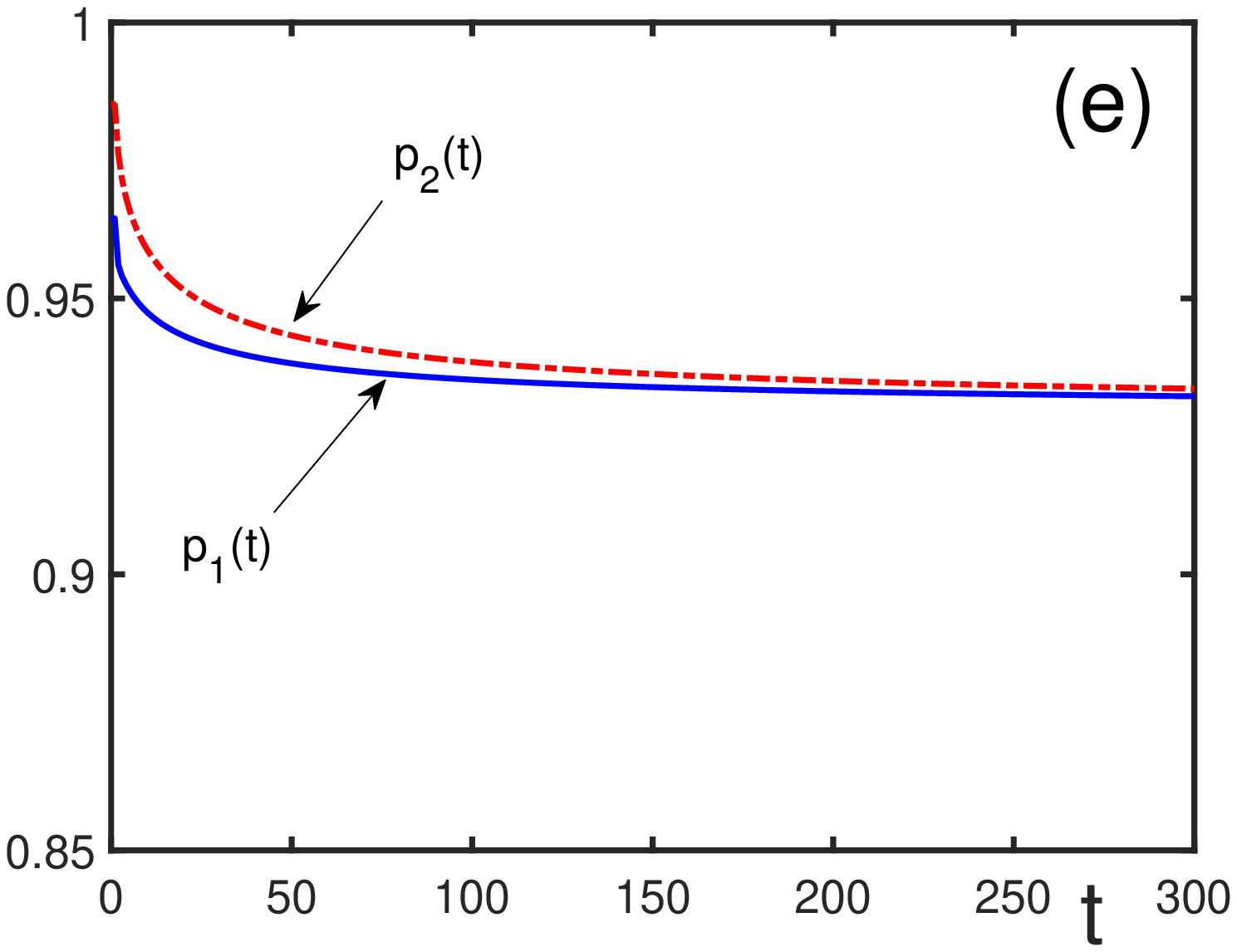} } }
\caption{Probabilities $p_1(t)$ (solid line) and $p_2(t)$ (dash-dotted line) 
for the initial conditions $f_1=0.1$, $f_2=0$, $q_1=0.899$, and $q_2=0.93$.
(a) $\ep_1=\ep_2=1$. The function $p_1(t)$ oscillates, while $p_2(t)$ tends 
monotonically to the fixed point $p_2^*=0.1$; 
(b) $\ep_1=1$ and $\ep_2=0.98$. Both functions $p_1(t)$ and $p_2(t)$ oscillate; 
(c) $\ep_1 = 1$ and $\ep_2 = 0.9$. The probabilities tend to the fixed points 
$p_1^*=0.6$ and $p_2^*=0.15$; 
(d) $\ep_1=1$ and $\ep_2=0.8$. Both functions converge to the same fixed point 
$p_1^*=p_2^*=0.93$; 
(e) $\ep_1=0.5$ and $\ep_1=0.8$. The functions converge monotonically to the 
consensual fixed point $p_1^* = p_2^* = 0.93$ .
}
\label{fig:Fig.7}
\end{figure}

\vskip 2mm
{\bf 8}. For the herding parameters on or close to the point $\ep_1+\ep_2=2$, there 
appear large chaotic fluctuations after rather long time from the beginning of the
process. First, only the probability $p_1(t)$ starts chaotically oscillating, while 
$p_2(t)$ smoothly tends to a fixed point. Diminishing the herding parameters results in 
the sudden chaotic oscillation of both functions. But the further decrease of the herding
parameters approaching the line $\ep_1+\ep_2=1$ , first, makes the oscillations periodic 
and then eliminates oscillations at al., so that both functions converge to the consensual 
fixed point (see Fig. 8).  

%Figure 8
\begin{figure}[ht]
\centerline{
\hbox{ \includegraphics[width=7.5cm]{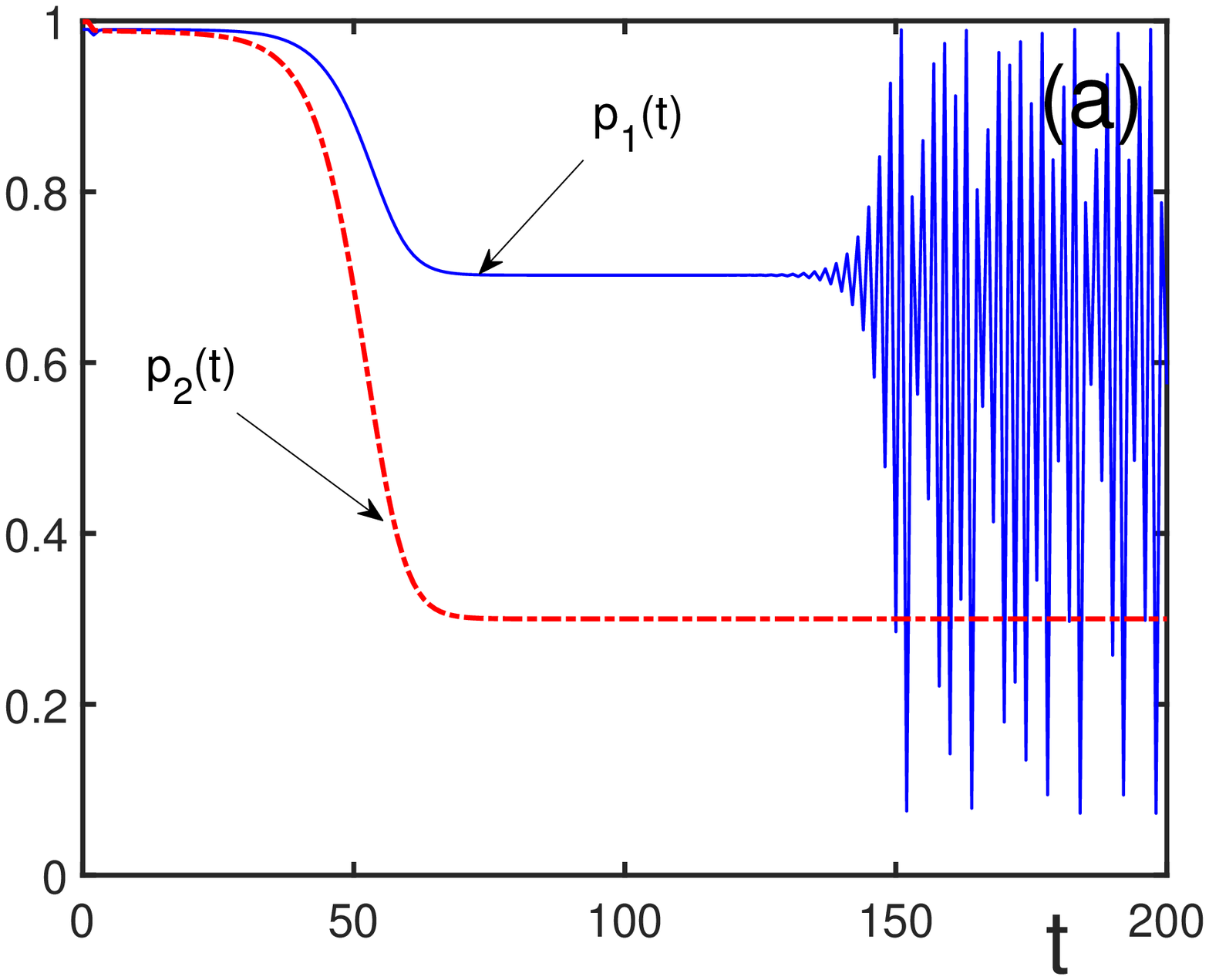} \hspace{1cm}
\includegraphics[width=7.5cm]{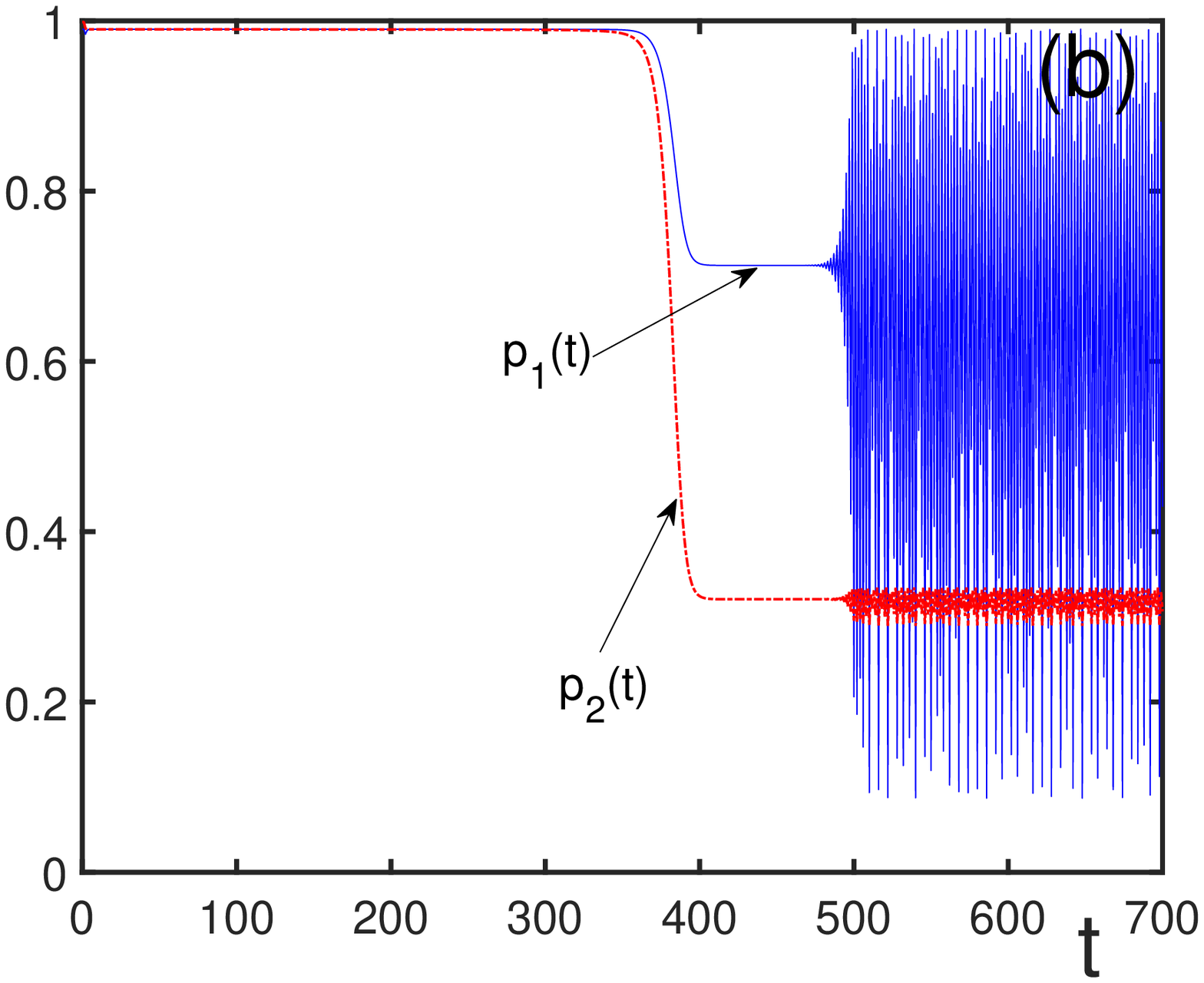}  } }
\vspace{12pt}
\centerline{
\hbox{ \includegraphics[width=7.5cm]{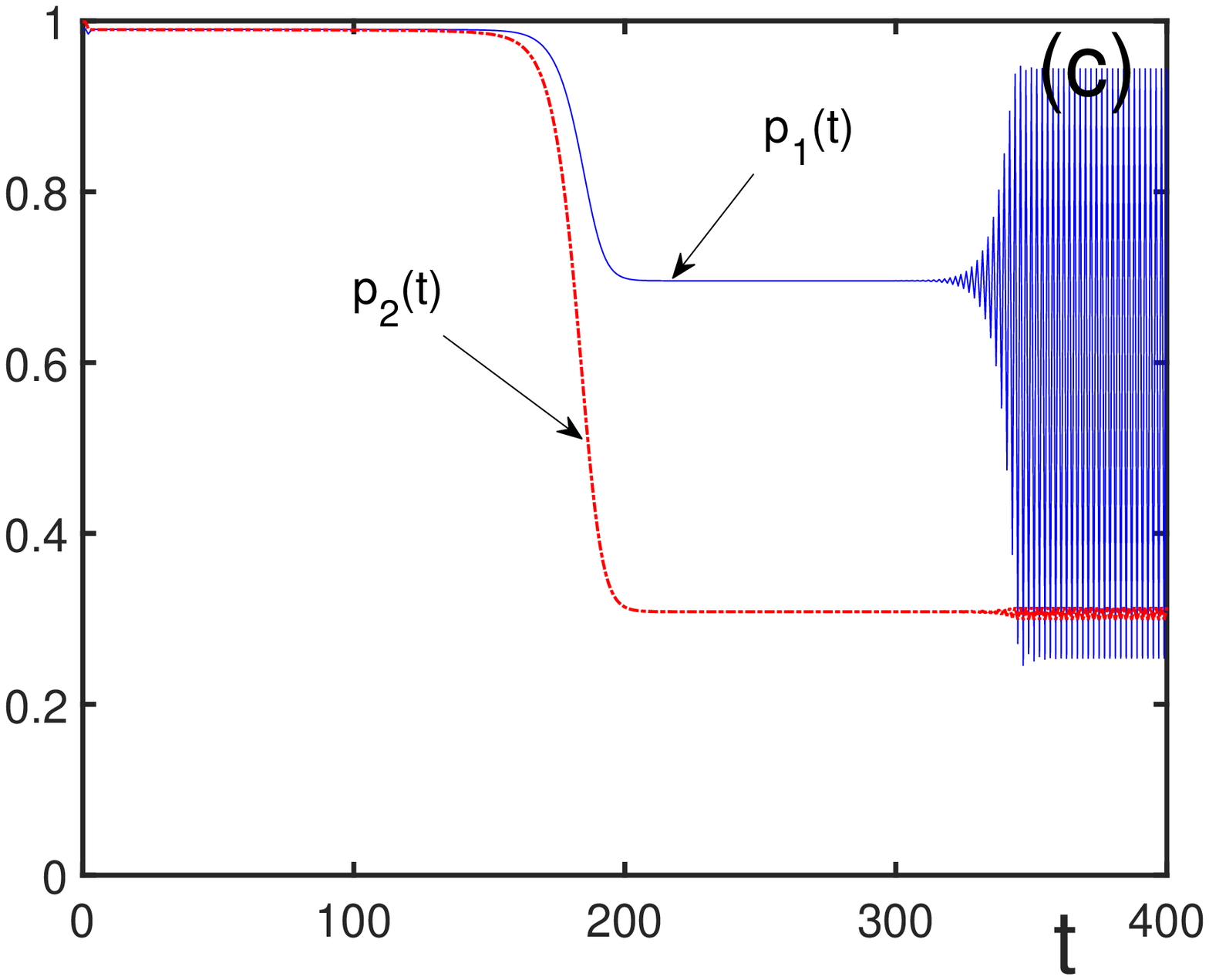} \hspace{1cm}
\includegraphics[width=7.5cm]{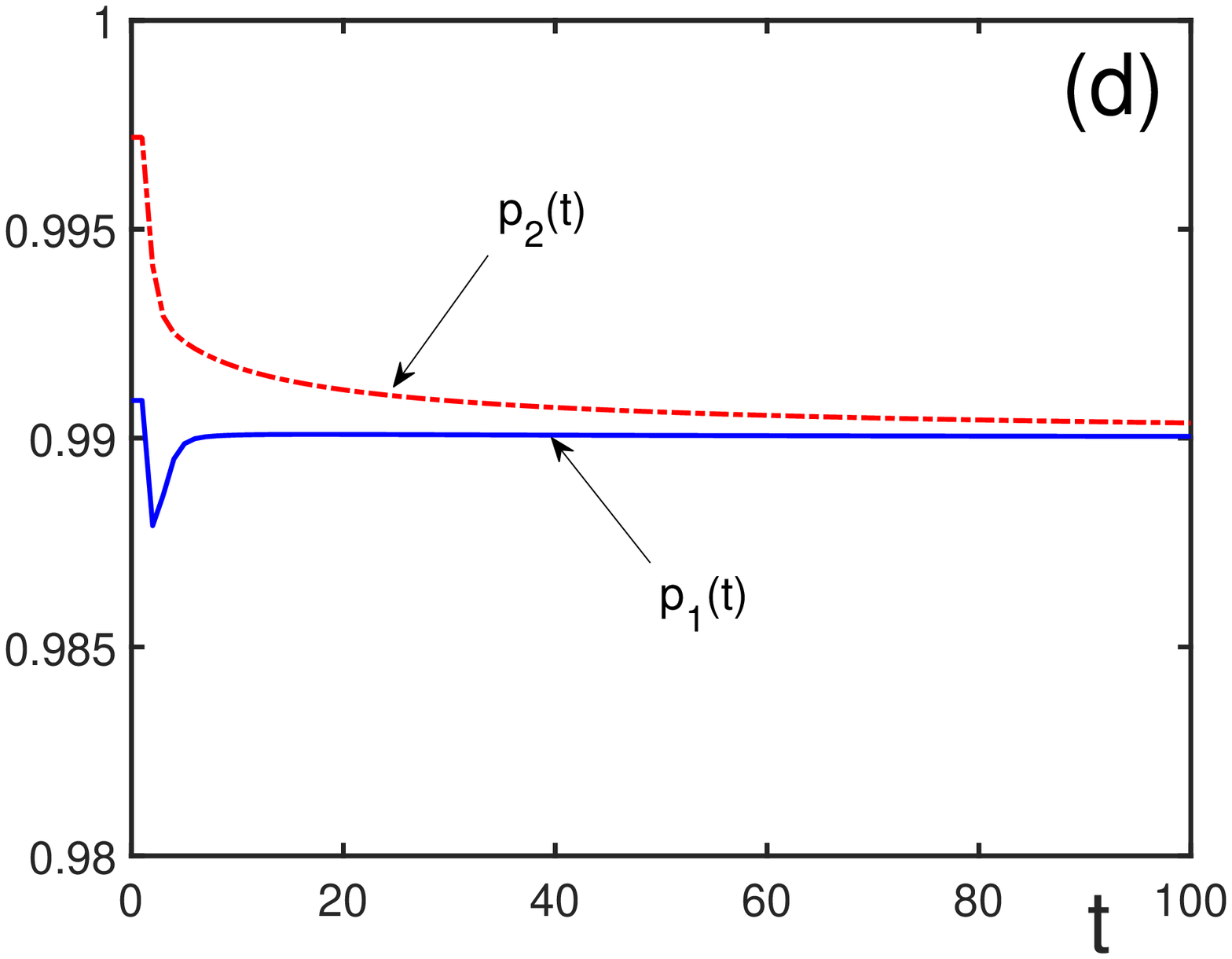} } }
\vspace{12pt}
\centerline{
\hbox{ \includegraphics[width=7.5cm]{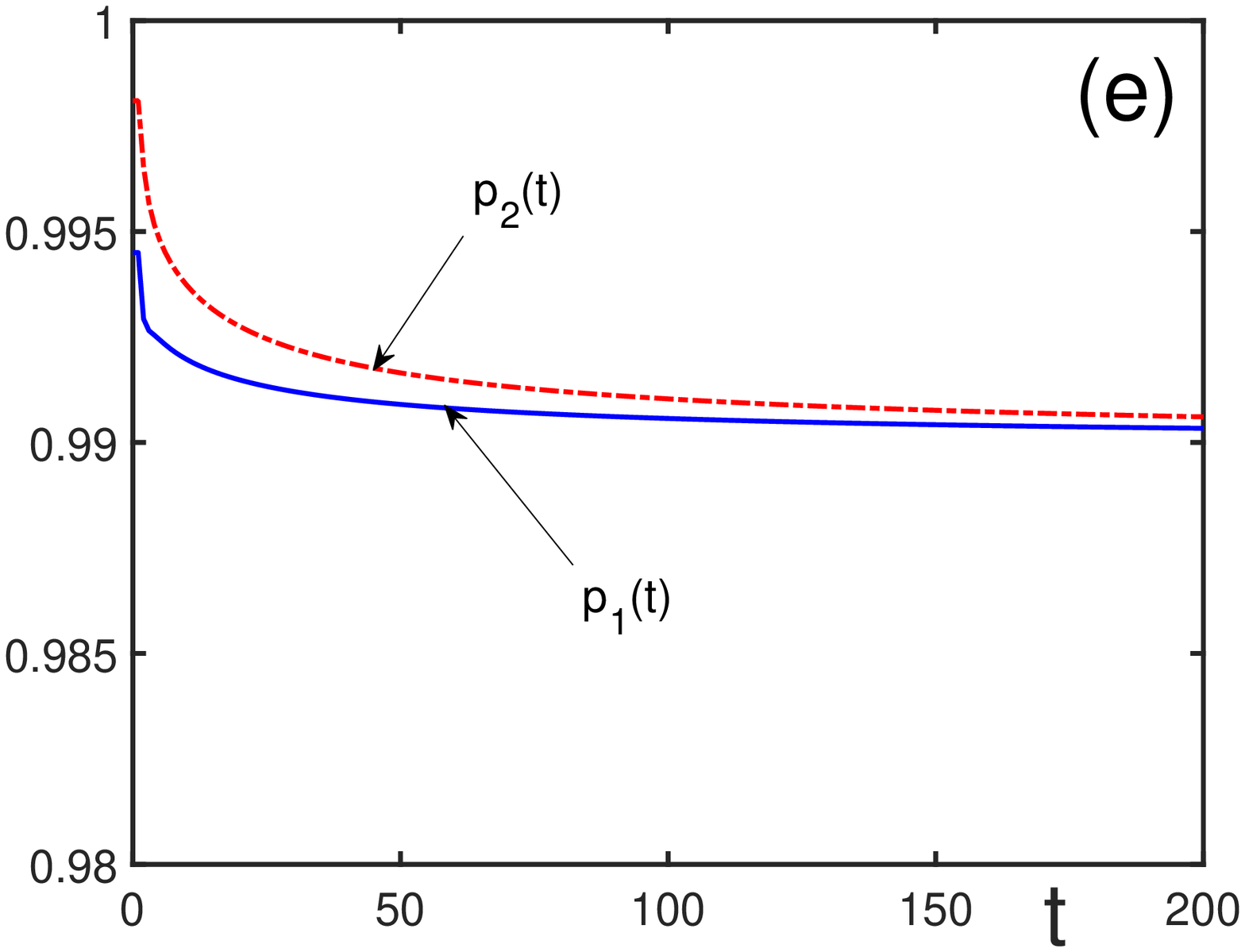} } }
\caption{Probabilities $p_1(t)$ (solid line) and $p_2(t)$ (dash-dotted line) 
for the initial conditions $f_1=0.3$, $f_2=0$, $q_1=0.699$, and $q_2=0.99$.
(a) $\ep_1=\ep_2=1$. The function $p_1(t)$, after quite long time of smooth 
behaviour, suddenly starts chaotically oscillating, but $p_2(t)$ goes 
monotonically to the fixed point $p_2^*=0.3$; 
(b) $\ep_1 = 1$ and $\ep_2=0.95$. Both functions $p_1(t)$ and $p_2(t)$, 
after rather long time of smooth behaviour, start chaotically oscillating; 
(c) $\ep_1=0.94$ and $\ep_2=0.98$. Both functions $p_1(t)$ and $p_2(t)$, 
after the initial smooth behaviour, start periodic oscillations; 
(d) $\ep_1=0.9$ and $\ep_2=0.8$. Oscillations are suppressed and the functions 
$p_1(t)$ and $p_2(t)$ converge to the consensual fixed point $p_1^*=p_2^*=0.99$; 
(e) $\ep_1=0.5$ and $\ep_2=0.9$. Both functions monotonically converge to the 
same fixed point $p_1^*=p_2^*=0.99$.
}
\label{fig:Fig.8}
\end{figure}

\vskip 2mm
{\bf 9}. At the point $\ep_1 + \ep_2 = 2$, the probability for the agents with
long-term memory begins chaotically oscillating from the beginning, while that for
agent with short-term memory monotonically tends to a fixed point. For smaller herding
parameters, first, both functions start chaotically oscillating, then chaotic oscillations
transfer to periodic fluctuations, and then both functions become monotonic, tending to 
fixed points. This behaviour is demonstrated in Fig. 9.

%Figure 9
\begin{figure}[ht]
\centerline{
\hbox{ \includegraphics[width=7.5cm]{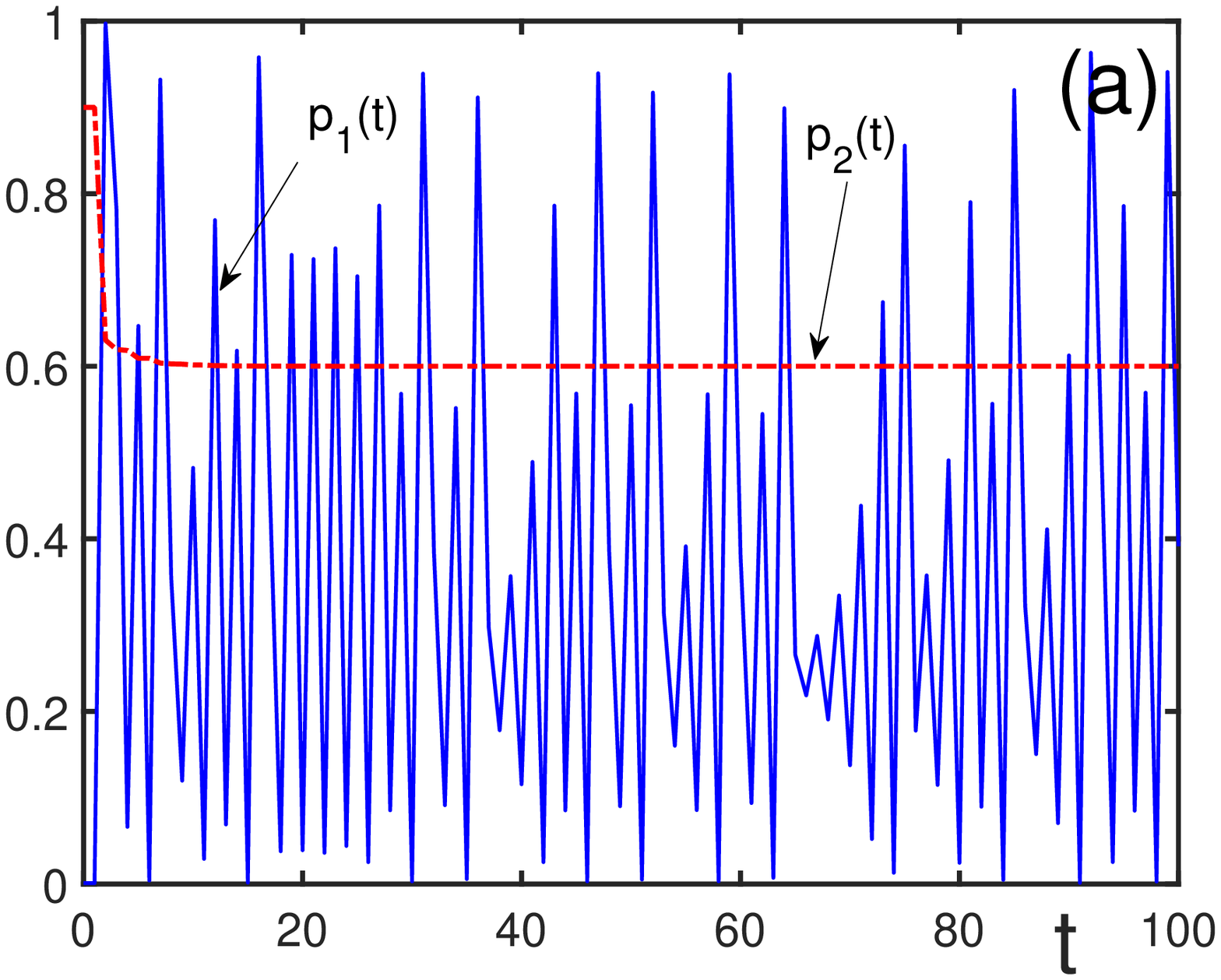} \hspace{1cm}
\includegraphics[width=7.5cm]{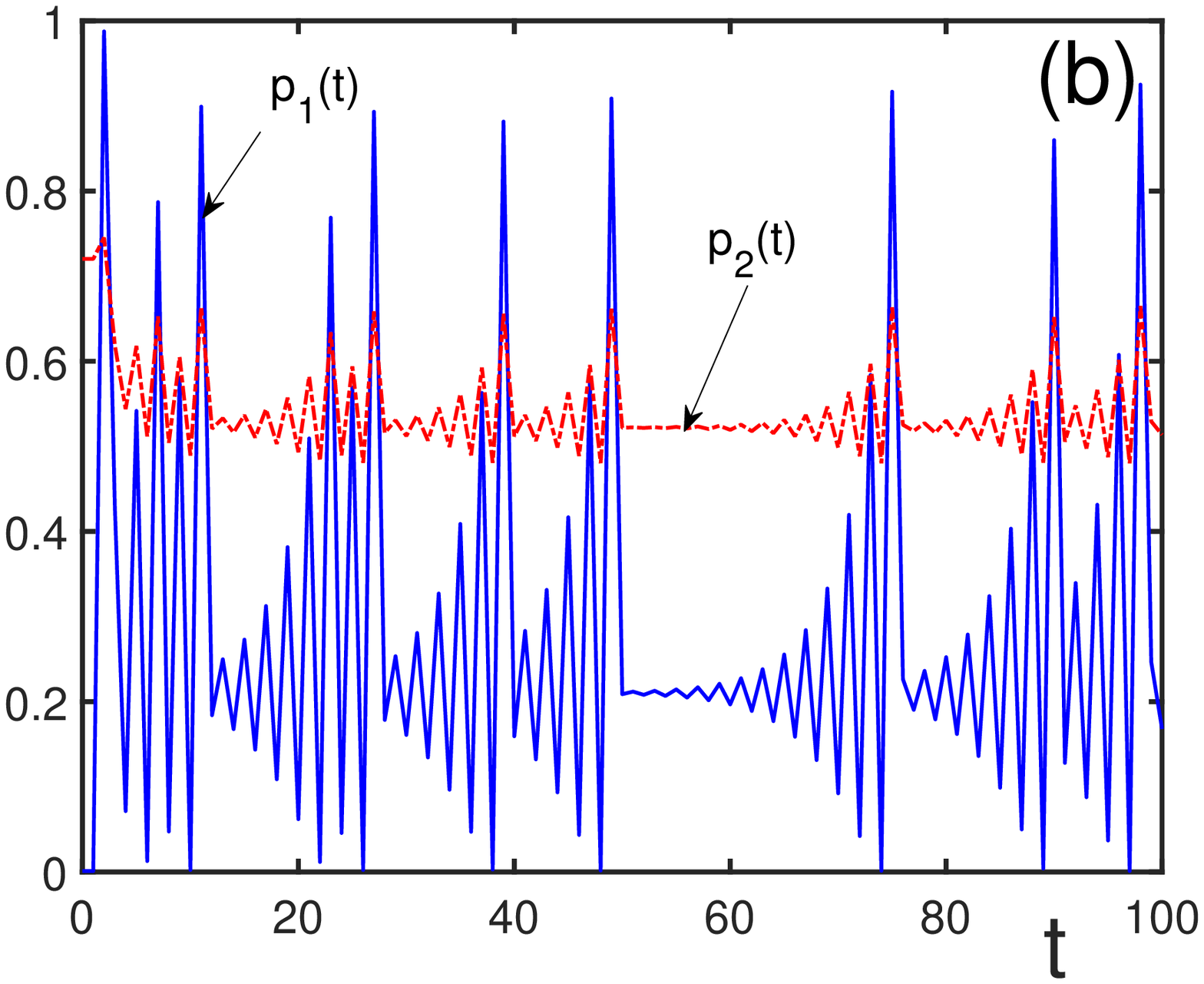}  } }
\vspace{12pt}
\centerline{
\hbox{ \includegraphics[width=7.5cm]{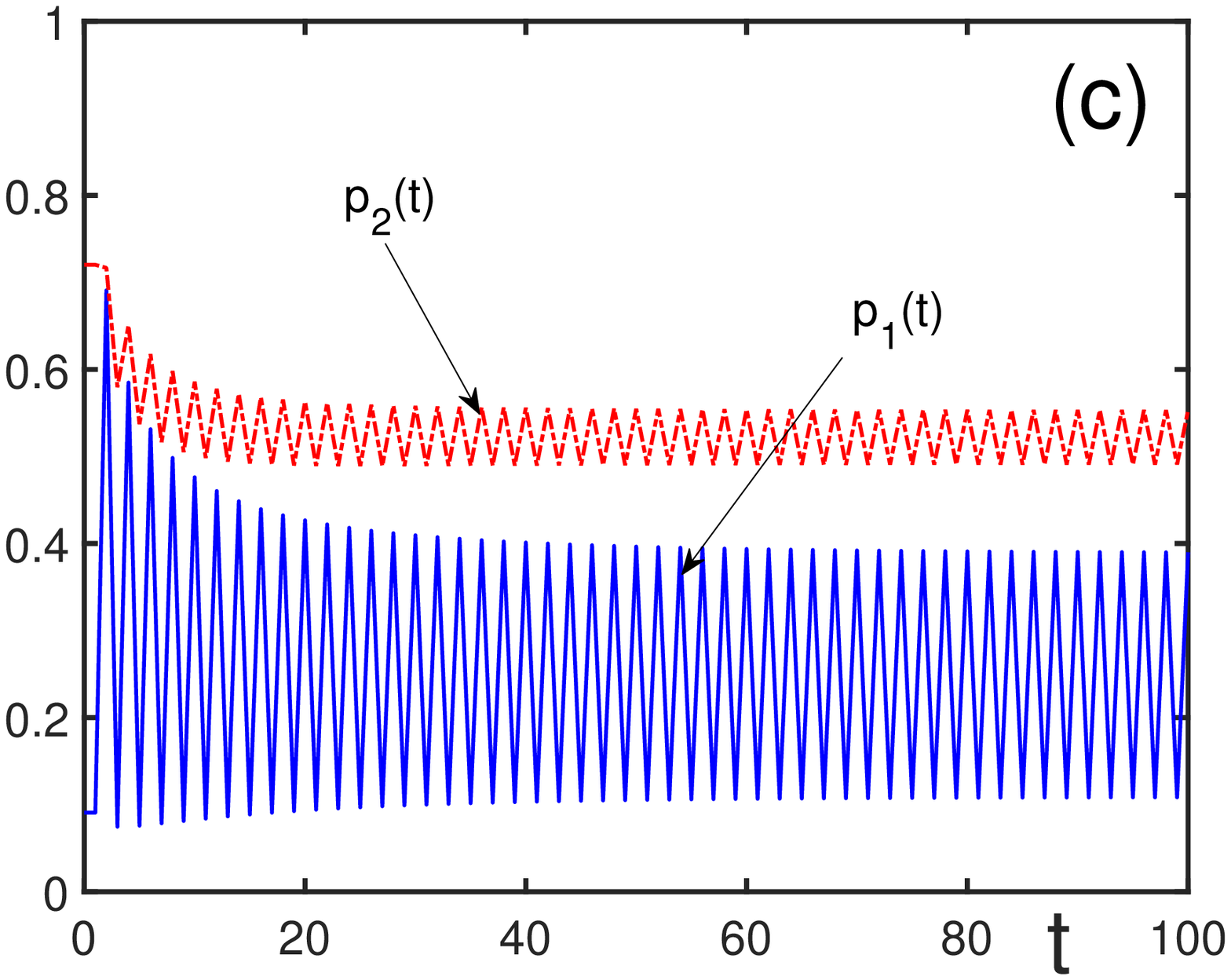} \hspace{1cm}
\includegraphics[width=7.5cm]{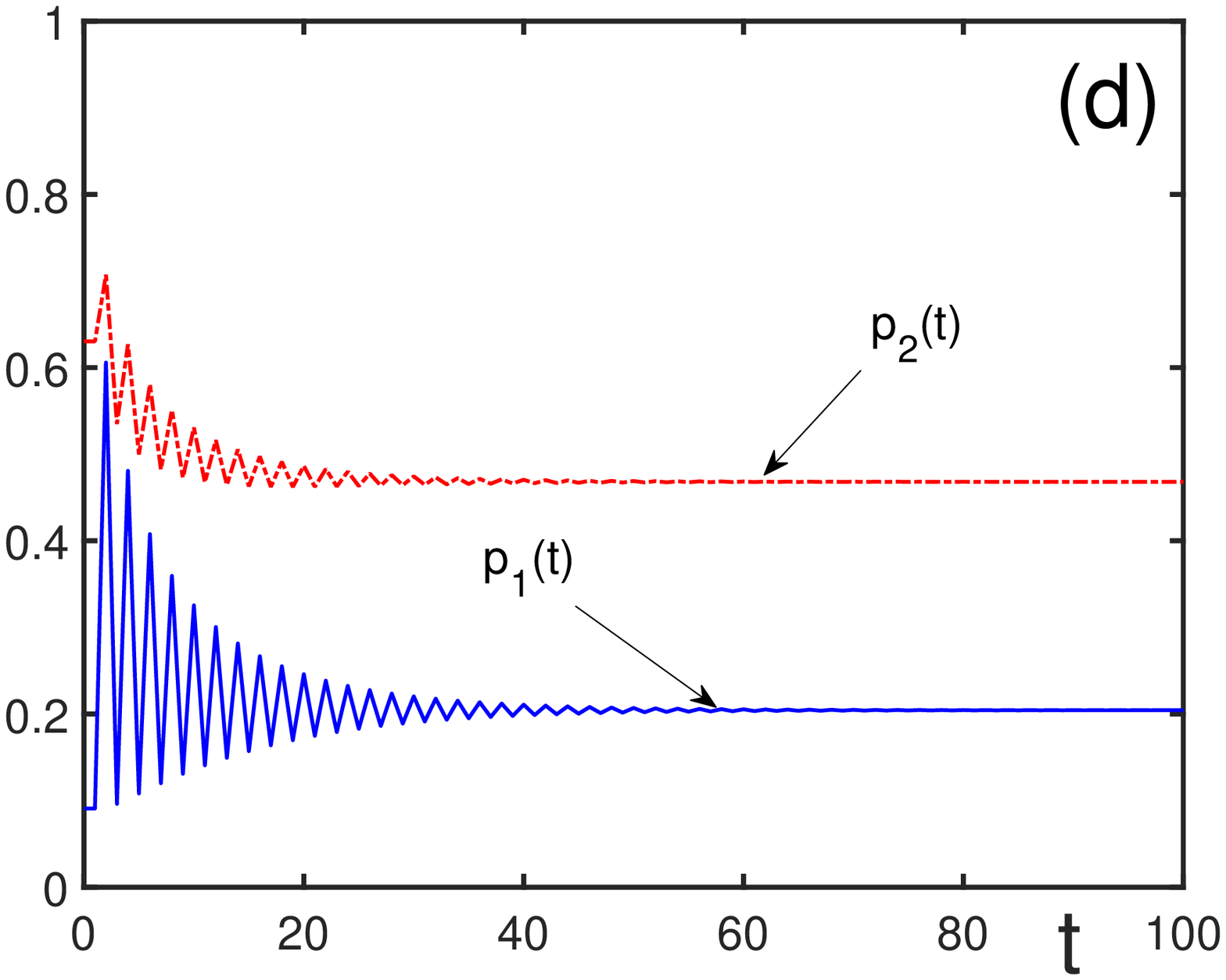} } }
\vspace{12pt}
\centerline{
\hbox{ \includegraphics[width=7.5cm]{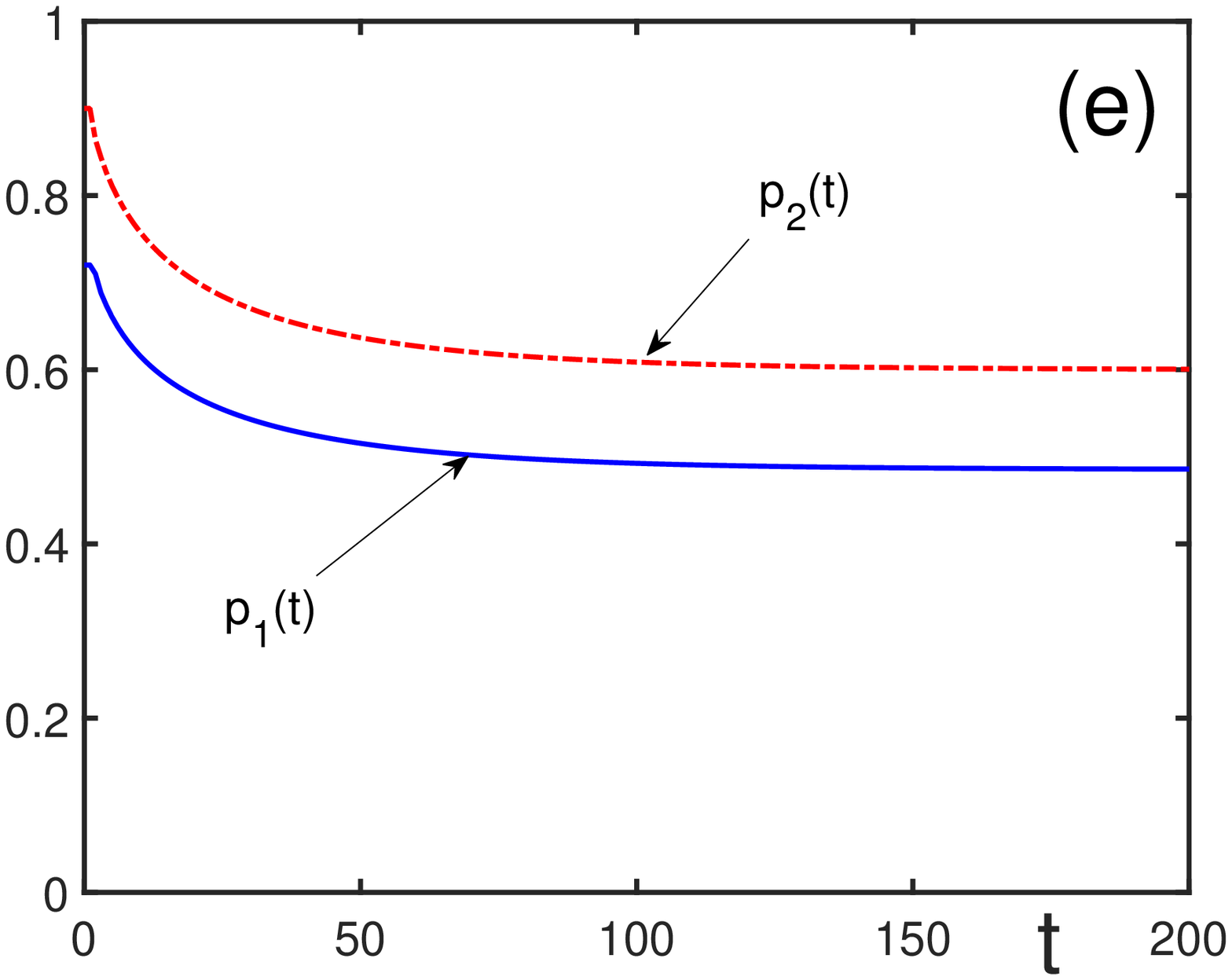} } }
\caption{Probabilities $p_1(t)$ (solid line) and $p_2(t)$ (dash-dotted line) 
for the initial conditions $f_1=0.6$, $f_2=1$, $q_1=0.3$, and $q_2=-0.999$.
(a) $\ep_1 = \ep_2 = 1$. The function $p_1(t)$ starts chaotically oscillating 
from the very beginning, while $p_2(t)$ monotonically tends to the fixed point 
$p_2^*=0.6$; 
(b) $\ep_1=1$ and $\ep_2=0.8$. Both functions $p_1(t)$ and $p_2(t)$ oscillate 
chaotically; 
(c) $\ep_1=0.9$ and $\ep_2=0.8$. Both functions $p_1(t)$ and $p_2(t)$ exhibit 
periodic oscillations; 
(d) $\ep_1=0.9$ and $\ep_2=0.7$. After the initial period of oscillations, both 
functions tend to the fixed points $p_1^*=0.204$ and $p_2^*=0.468$; 
(e) $\ep_1=0.2$ and $\ep_2=1$. Both functions monotonically converge to the 
fixed points $p_1^*=0.485$ and $p_2^*=0.6$.
}
\label{fig:Fig.9}
\end{figure}

\vskip 2mm
{\bf 10}. One more example of chaotic fluctuations for large herding parameters, which, 
first, become periodic and then are suppressed when moving from the point $\ep_1+\ep_2=2$ 
to the line $\ep_1+\ep_2=1$. The transformation of dynamic regimes, when reducing the 
herding parameters, is shown in Fig. 10.

%Figure 10
\begin{figure}[ht]
\centerline{
\hbox{ \includegraphics[width=7.5cm]{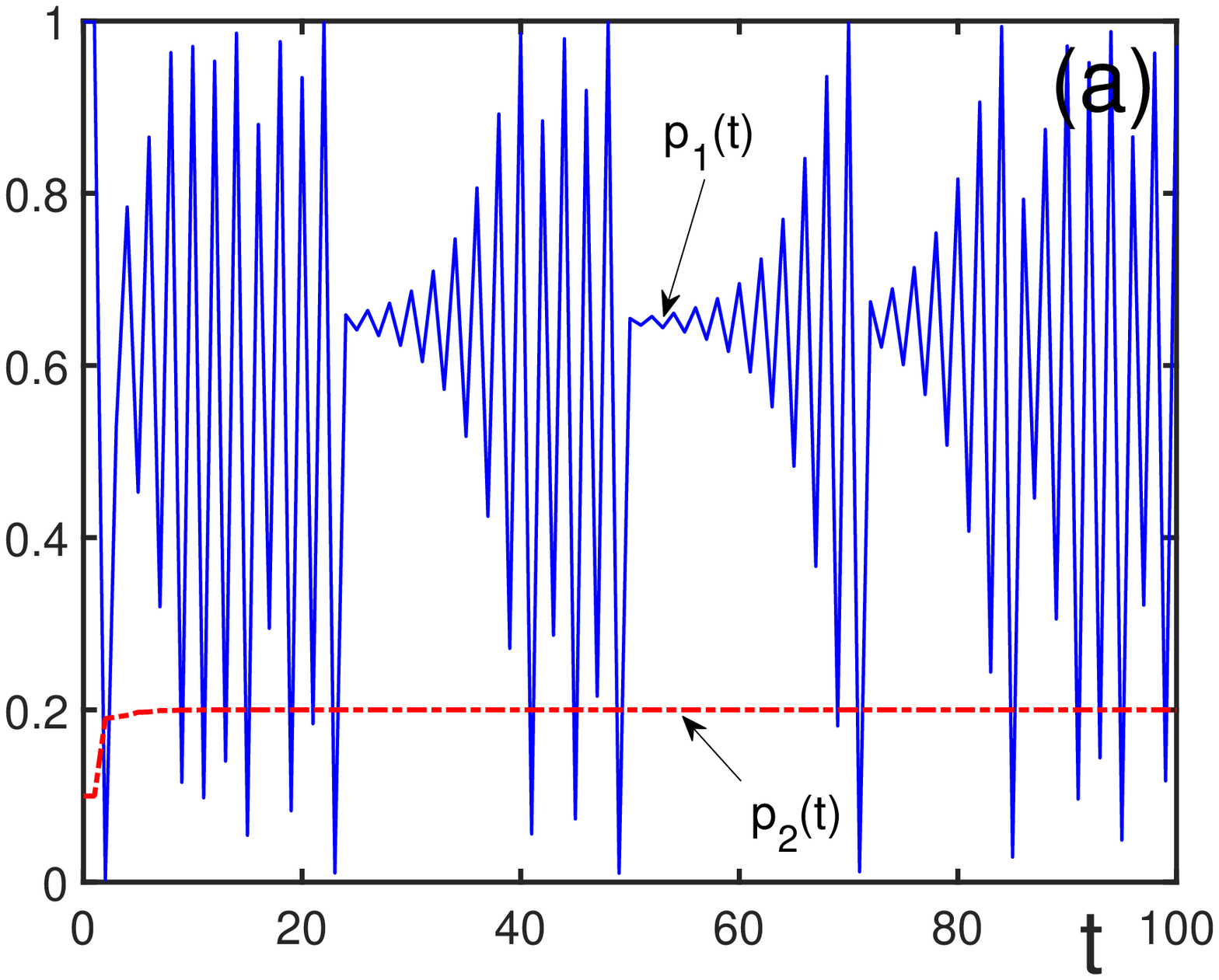} \hspace{1cm}
\includegraphics[width=7.5cm]{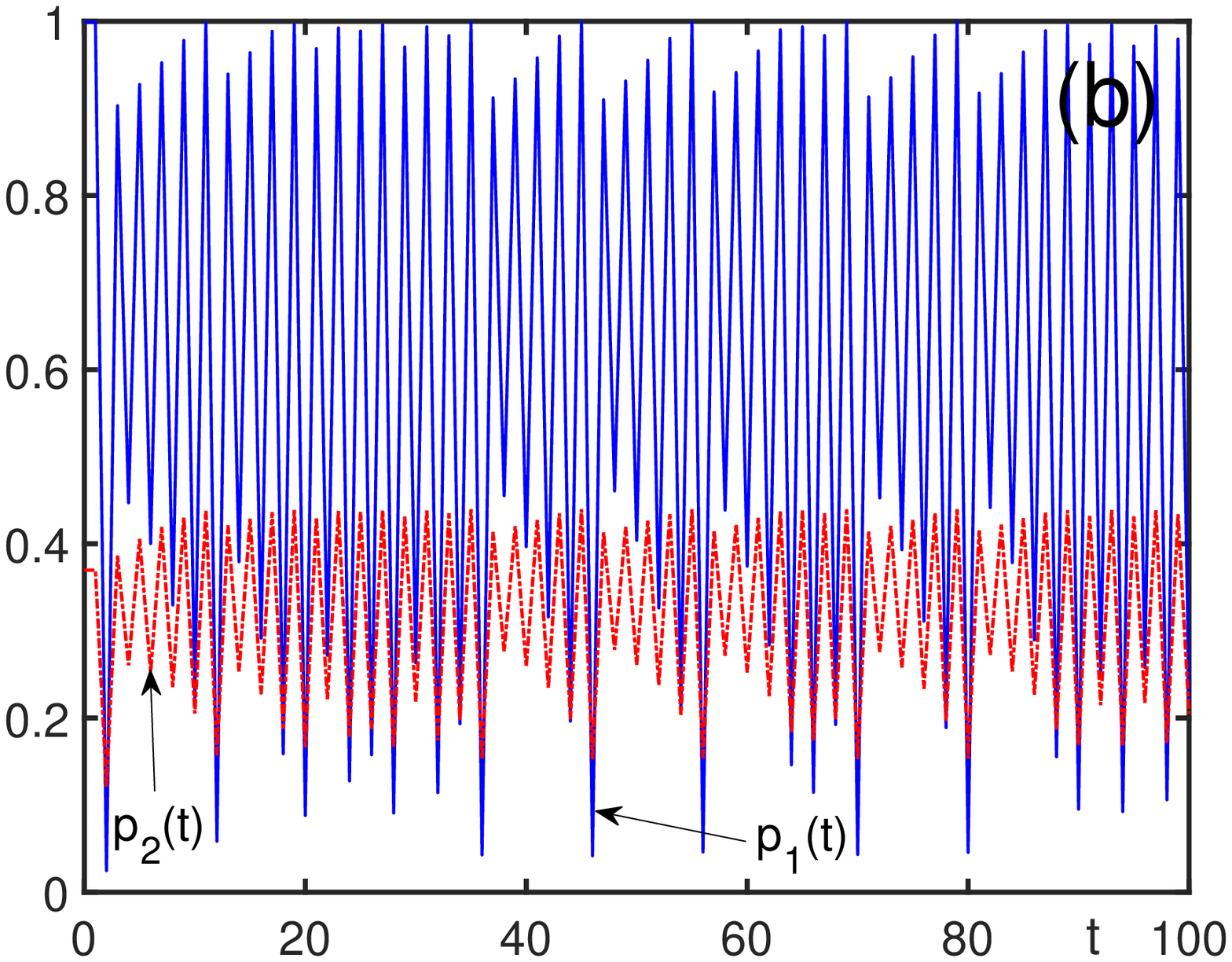}  } }
\vspace{12pt}
\centerline{
\hbox{ \includegraphics[width=7.5cm]{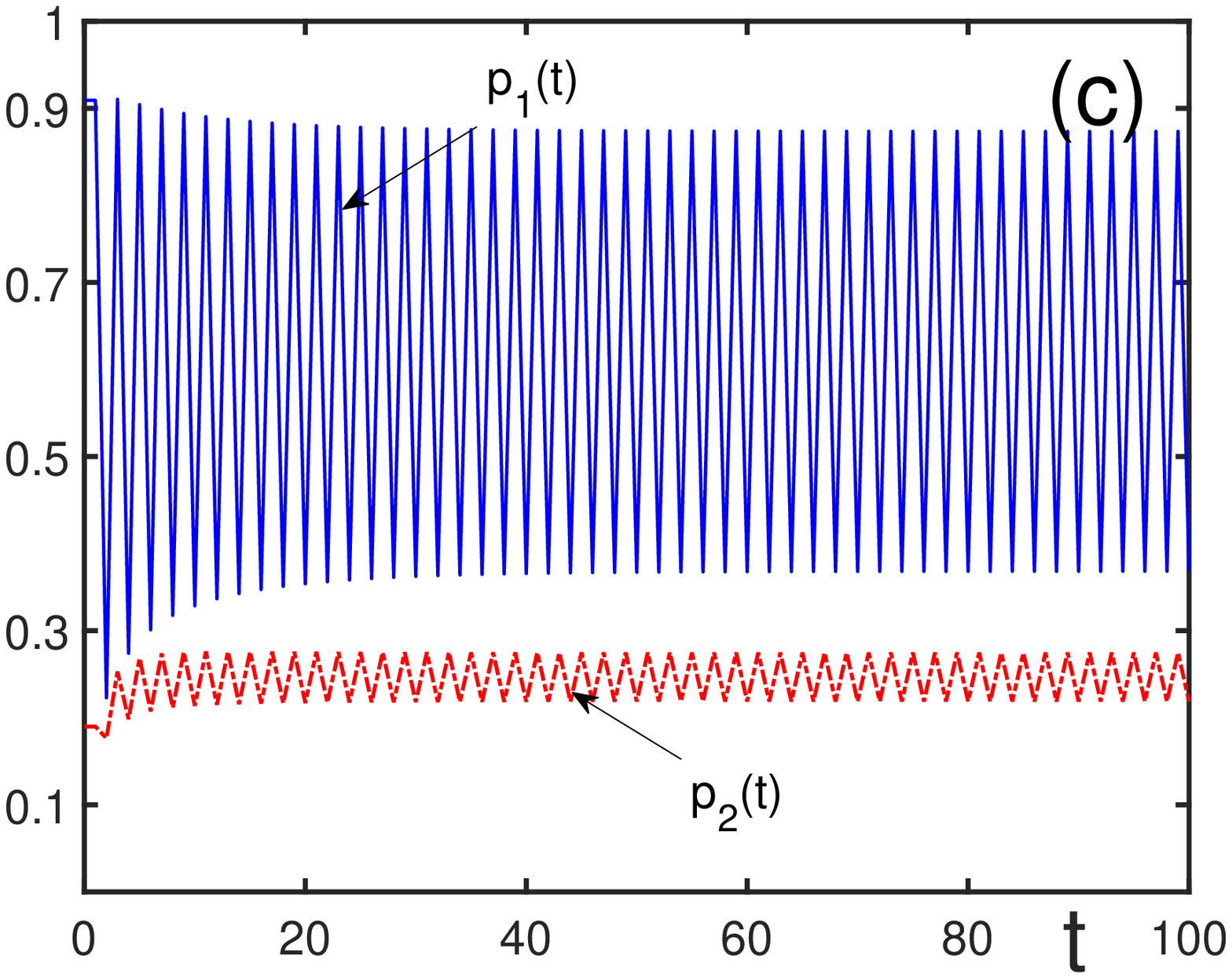} \hspace{1cm}
\includegraphics[width=7.5cm]{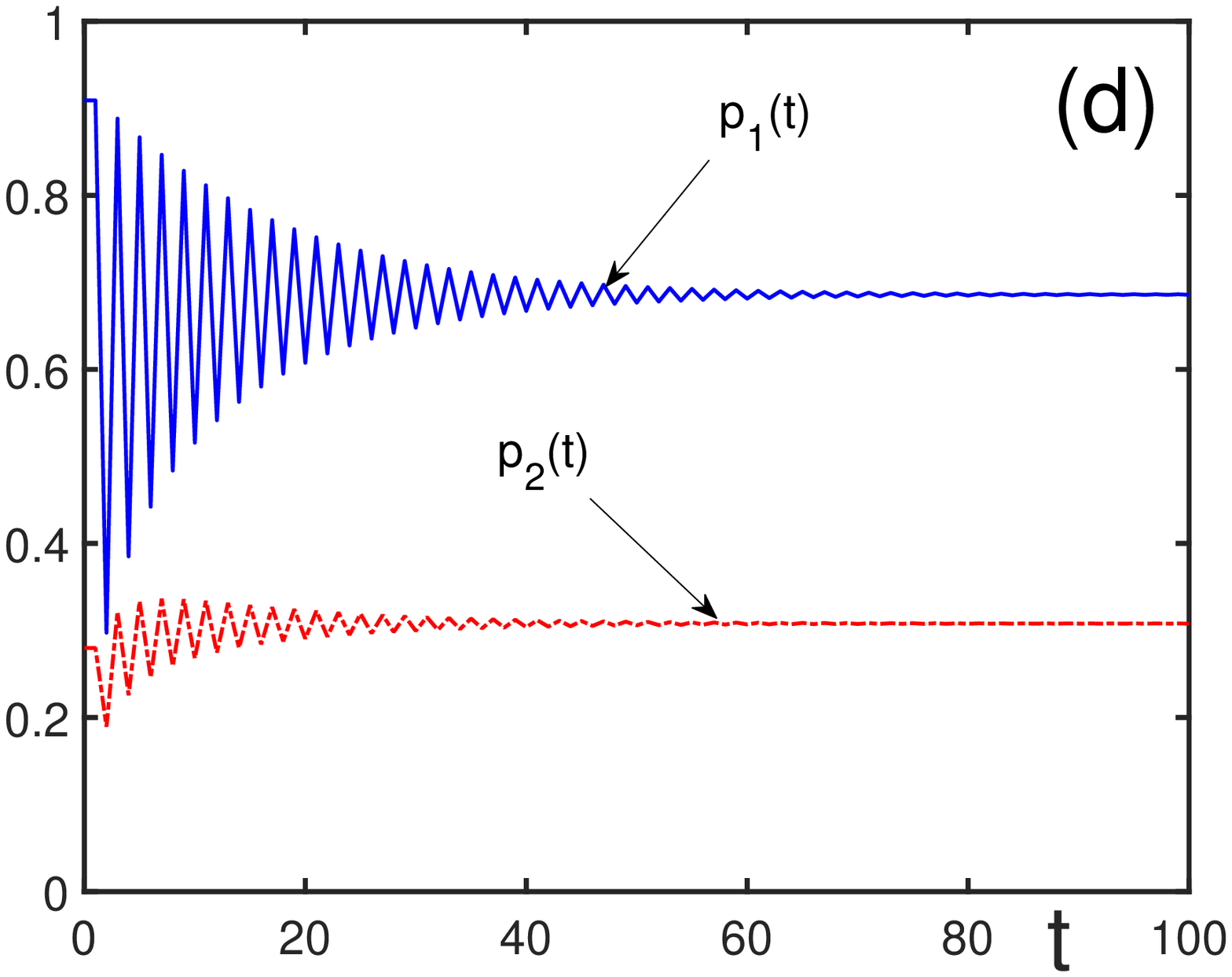} } }
\vspace{12pt}
\centerline{
\hbox{ \includegraphics[width=7.5cm]{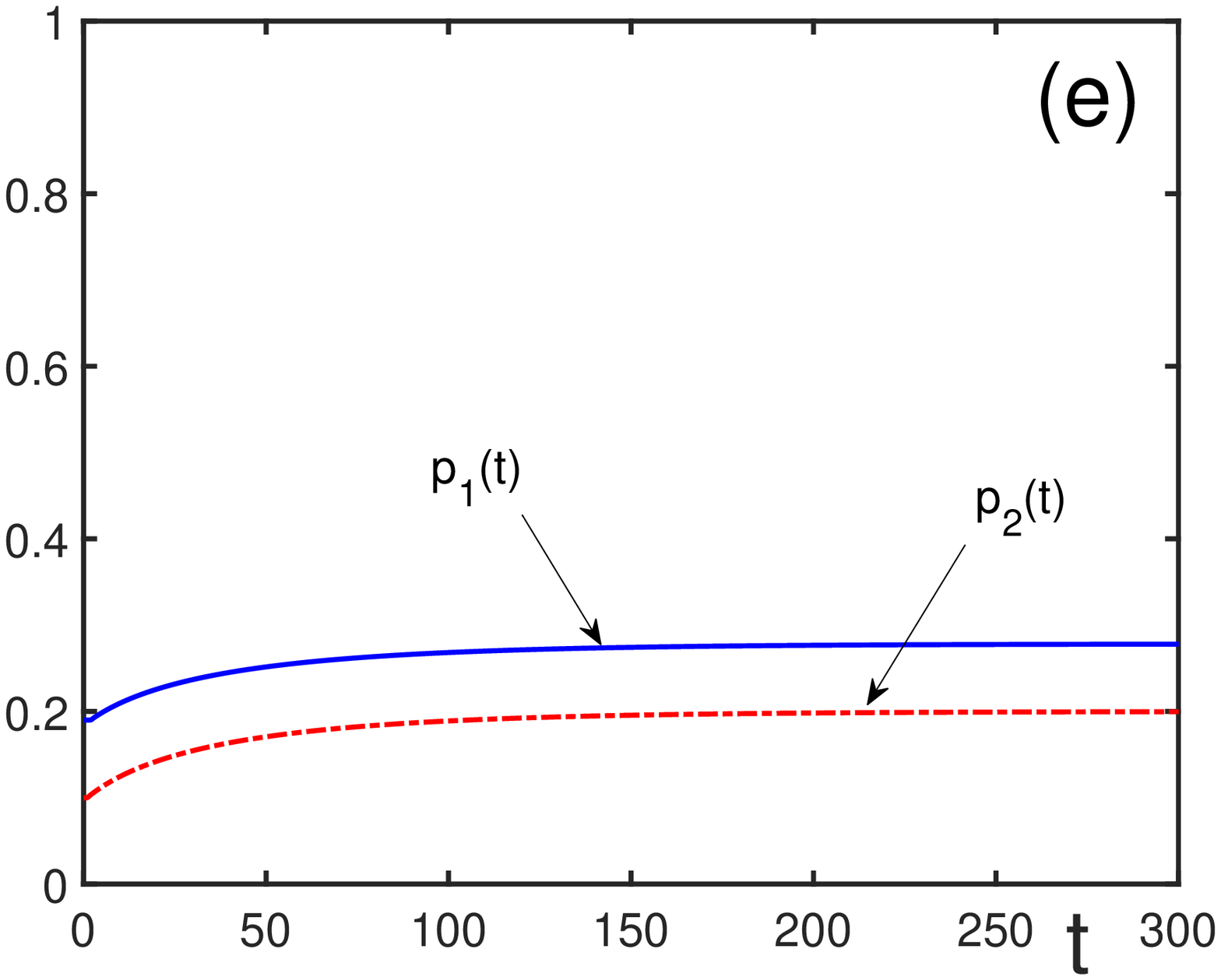} } }
\caption{Probabilities $p_1(t)$ (solid line) and $p_2(t)$ (dash-dotted line) 
for the initial conditions $f_1=0.2$, $f_2=0$, $q_1=-0.1$, and $q_2=0.999$.
(a) $\ep_1=\ep_2=1$. The function $p_1(t)$ experiences chaotic oscillations, 
while $p_2(t)$ monotonically tends to the fixed point $p_2^*=0.2$; 
(b) $\ep_1=1$ and $\ep_2=0.7$. Both functions $p_1(t)$ and $p_2(t)$ chaotically 
oscillate; 
(c) $\ep_1=\ep_2=0.9$. Both functions $p_1(t)$ and $p_2(t)$ demonstrate periodic 
oscillations; 
(d) $\ep_1=0.9$ and $\ep_2=0.8$. After initial fluctuations, the functions tend 
to the fixed points $p_1^*=0.685$ and $p_2^*=0.308$; 
e) $\ep_1=0.1$ and $\ep_2=1$. Both functions monotonically tend to the fixed 
points $p_1^*=0.278$ and $p_2^*=0.2$.
}
\label{fig:Fig.10}
\end{figure}

\section{Conclusion}

We developed a model of a society composed of intelligent agents making decisions by
choosing between several alternatives. The choice is based of three attributes, utility,
attraction, and herding. By varying the herding parameters there appear different regimes
of decision making dynamics. A detailed investigation of possible regimes is done for
the time dependence of probabilities of choosing one of two alternatives. Two groups of 
decision makers are considered, one group consisting of agents with long-term memory
and the other group composed of agents with short-term memory. 

The most interesting observation is the sudden appearance of oscillations that can be 
either periodic or chaotic. They can arise either from the beginning of the decision
process or after rather long time of smooth dynamics. These suddenly arising waves
are self organized, since the society is a closed system having no external 
perturbations. The sudden growth of the self-excited waves is an interesting example
of the appearance of periodic and chaotic dynamics in a closed system. The herding 
effect smooths the dynamics and even can suppress the self-excited waves. 

The aim of the paper is the analysis of admissible dynamic regimes in a complex society. 
Qualitatively similar dynamics regimes can occur in real social systems. As illustrations,
it is possible to adduce several dynamic trends from Google analogous to the evolution 
curves in our figures. Thus, Fig. 11 demonstrates the curve rising to a limiting value,
with small random fluctuations. Figure 12 depicts the decreasing tendency to a limit.   
In Fig. 13, we see almost periodic motion. And Fig. 14 displays a kind of chaotic behavior.
It looks that all types of behavior exhibited by the model we considered can happen in
realistic social systems.

%Figure 11
\begin{figure}[ht]
\centerline{
\includegraphics[width=10cm]{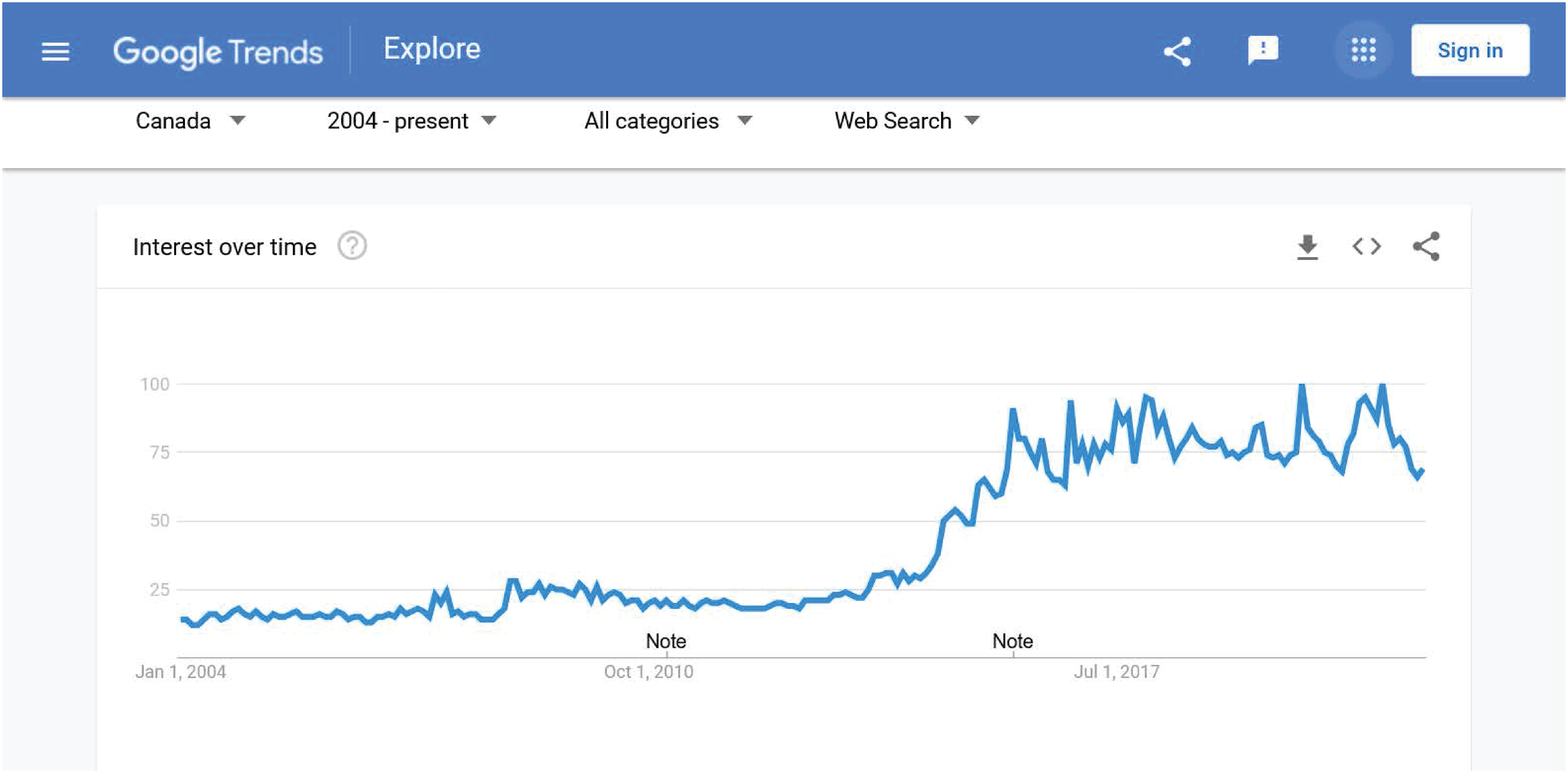}  }
\caption{Interest to USD in Canada over time from 2004 up to now. (Increases to 
its limit).} 
\label{fig:Fig.11}
\end{figure}

%Figure 12
\begin{figure}[ht]
\centerline{
\includegraphics[width=10cm]{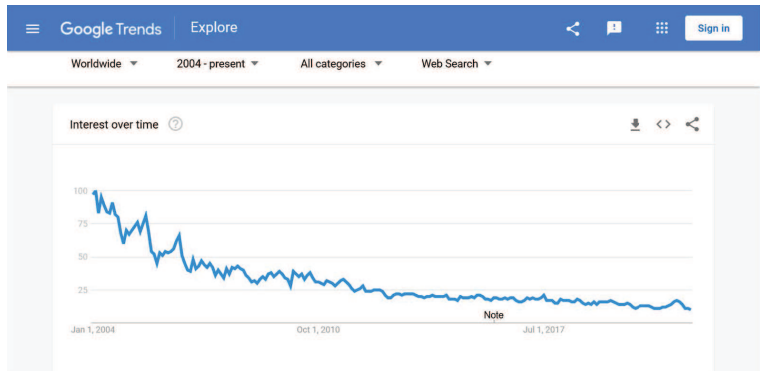}  }
\caption{Interest to car prices worldwide over time from 2004 up to now.
(Decreases to its limit).} 
\label{fig:Fig.12}
\end{figure}

%Figure 13
\begin{figure}[ht]
\centerline{
\includegraphics[width=10cm]{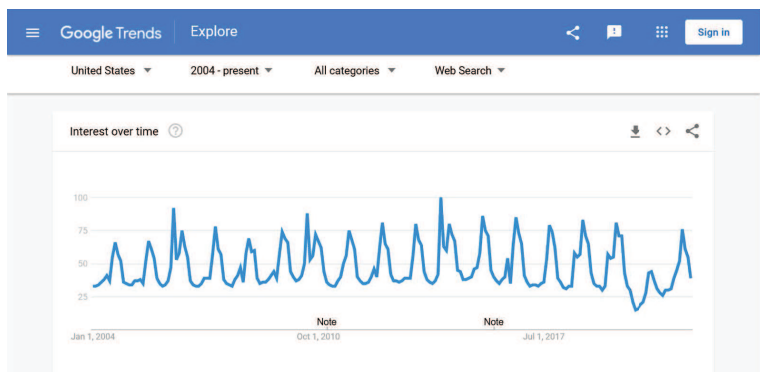}  }
\caption{Interest to football game in USA over time from 2004 up to now.
(Oscillations).} 
\label{fig:Fig.13}
\end{figure}

%Figure 14
\begin{figure}[ht]
\centerline{
\includegraphics[width=10cm]{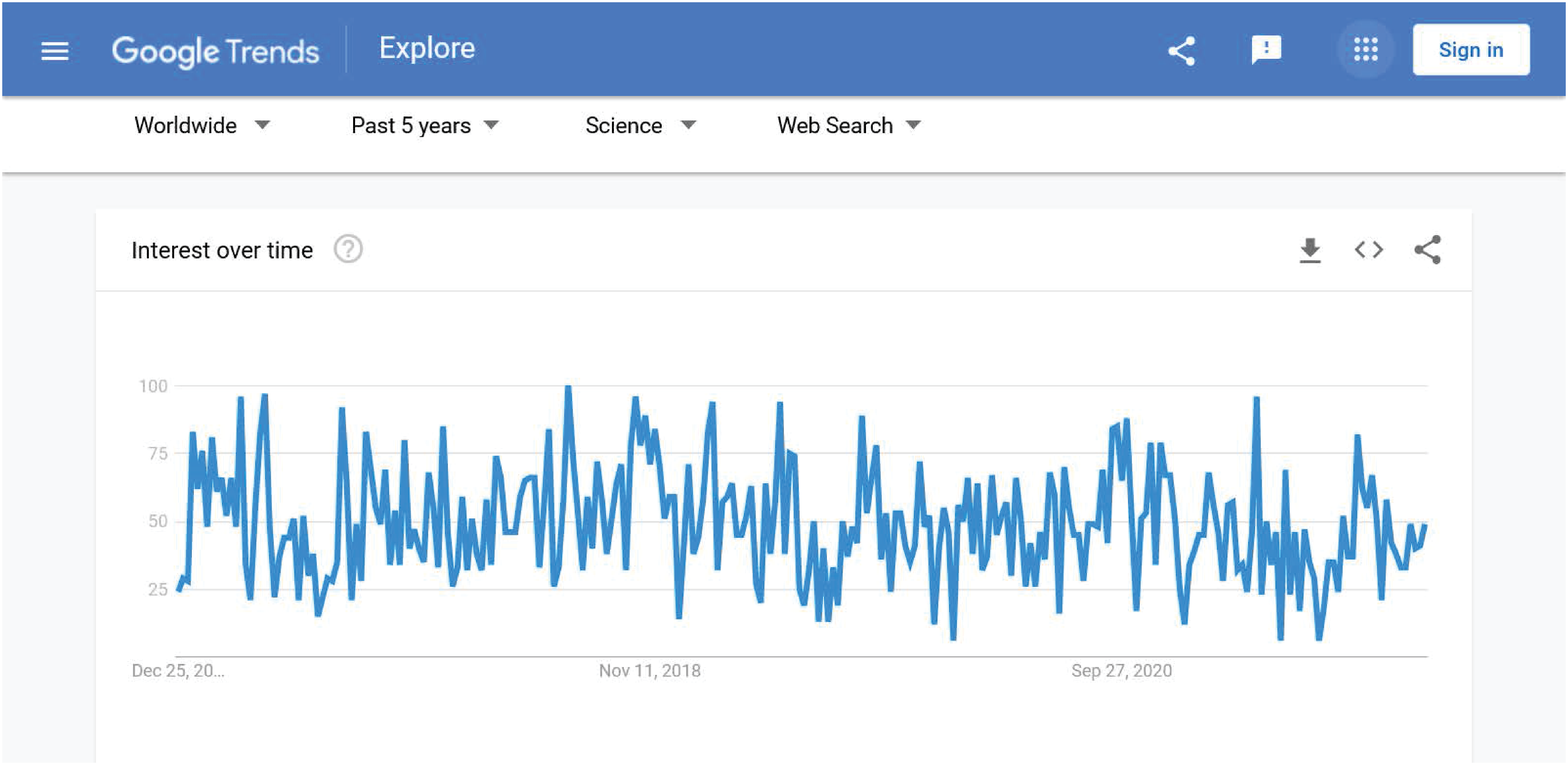}  }
\caption{Interest to liberalism as a political philosophy in Science category
over time for 5 year up to now. (Chaotic behavior).} 
\label{fig:Fig.14}
\end{figure}

\vskip 2mm
{\bf CRediT authorship contribution statement}
\vskip 2mm

V.I. Yukalov and E.P. Yukalova equally contributed to the paper:
Concept, Design, Analysis, Writing, or revision of the manuscript.

\vskip 2mm
{\bf Declaration of competing interest}
\vskip 2mm

The authors declare that they have no known competing financial
interests or personal relationships that could have appeared
to influence the work reported in this paper.


\vskip 2cm
\begin{thebibliography}{99}
\bibitem{Parsons_1}
T. Parsons,  
The Social System, 
Taylor and Francis, London, 2005. 

\bibitem{Perc_2}
M. Perc, J. G\'{o}mez-Garde\~{n}es, A. Szolnoki, L.M. Floria, Y. Moreno, 
Evolutionary dynamics of group interactions on structured populations: a review, 
J. Roy. Soc. Interface 10 (2013) 20120997.

\bibitem{Perc_3}
M. Perc, J.J. Jordan, D.G. Rand, Z. Wang, S. Boccaletti, A. Szolnoki,
Statistical physics of human cooperation,
Phys. Rep. 687 (2017) 1--51.

\bibitem{Jusup_3}
M. Jusup, P. Holme, K. Kanazawa, M. Takayasu, I. Romic, Z. Wang, S. Gecek, T. Lipic, 
B. Podobnik, L. Wang, W. Luo, T. Klanjscek, J. Fan, S. Boccaletti, M. Perc,
Social physics,
arXiv:2110.01866 (2021).

\bibitem{Scott_4}
J.P. Scott, 
Social Network Analysis, 
Sage, Thousand Oaks, 2000.

\bibitem{Albert_5}
R. Albert, A.L. Barabasi, 
Statistical mechanics of complex networks, 
Rev. Mod. Phys. 74 (2002) 47–98.

\bibitem{Boccaletti_6}
S. Boccaletti, V. Latora, Y. Moreno, M. Chavez, D.U. Hwanga,
Complex networks: Structure and dynamics,
Phys. Rep. 424 (2006) 175--308.

\bibitem{Meter_7}
R. Van Meter, 
Quantum Networking, Hoboken, Wiley, 2014.

\bibitem{Kennett_8}
D.Y. Kenett, M. Perc, S. Boccaletti,
Networks of networks: An introduction,
Chaos Solit. Fract. 80 (2015) 1--6.

\bibitem{Mata_9}
A.S. da Mata,
Complex networks: A mini review,
Braz. J. Phys. 50 (2020) 658--672.

\bibitem{Nilsson_10}
N. Nilsson, 
Artificial Intelligence: A New Synthesis, 
Morgan Kaufmann, San Francisco, 1998.

\bibitem{Poole_11}
D. Poole, A. Mackworth, R. Goebel, 
Computational Intelligence: A Logical Approach, 
Oxford University, New York, 1998.

\bibitem{Luger_12}
G.F. Luger, W.A. Stubblefield, 
Artificial Intelligence: Structures and Strategies for Complex Problem Solving, 
Benjamin Cummings, Redwood City, 2004.

\bibitem{Rich_13}
E. Rich, K. Knight, S.B. Nair, 
Artificial Intelligence, 
McGraw Hill, New Delhi, 2009.

\bibitem{Russell_14}
S.J. Russell, P. Norvig, 
Artificial Intelligence: A Modern Approach, Hoboken, Pearson, 2021.

\bibitem{Yukalov_15}
V.I. Yukalov, D. Sornette,
Quantum decision theory as quantum theory of measurement, 
Phys. Lett. A 372 (2008) 6867--6871.

\bibitem{Yukalov_16}
V.I. Yukalov, D. Sornette, 
Scheme of thinking quantum systems, 
Laser Phys. Lett. 6 (2009) 833--839.

\bibitem{Yukalov_17}
V.I. Yukalov, D. Sornette, 
Physics of risk and uncertainty in quantum decision making,
Eur. Phys. J. B 71 (2009) 533--548.

\bibitem{Yukalov_18}
V.I. Yukalov, D. Sornette,  
Decision theory with prospect interference and entanglement,
Theory Decis. 70 (2011) 283--328.

\bibitem{Yukalov_19}
V.I. Yukalov, D. Sornette,
Quantitative predictions in quantum decision theory,
IEEE Trans. Syst. Man Cybern. Syst. 48 (2018) 366--381. 

\bibitem{Yukalov_20}
V.I. Yukalov, 
Evolutionary processes in quantum decision theory,
Entropy 22 (2020) 681.

\bibitem{Yukalov_21}
V.I. Yukalov, 
Tossing quantum coins and dice,
Laser Phys. 31 (2021) 055201.

\bibitem{Yukalov_YYS}
V.I. Yukalov, E.P. Yukalova, D. Sornette,
Role of collective information in networks of quantum operating agents,
Physica A 598 (2022) 127365.

\bibitem{Yukalov_22}
V.I. Yukalov,
Quantification of emotions in decision making,
Soft Comput. 26 (2021) 2419--2436. 

\bibitem{Yukalov_23}
V.I. Yukalov,
A resolution of St. Petersburg paradox,
J. Math. Econ. 97 (2021) 102537.

\bibitem{Fishburn_24}
P.C. Fishburn,   
Decision and Value Theory, 
Wiley, New York, 1964.

\bibitem{Keeney_25}
R.L. Keeney, H. Raiffa,  
Decisions with Multiple Objectives: Preferences and Value Tradeoffs, 
Wiley, New York, 1976.

\bibitem{Frederick_26}
S. Frederick, G. Loewenstein, T. O'Donoghue, 
Time discounting and time preference: A critical review, 
J. Econ. Liter. 40 (2002) 351--401.

\bibitem{Yukalov_27}
V.I. Yukalov, E.P. Yukalova, D. Sornette, 
Information processing by networks of quantum decision makers, 
Physica A 492 (2018) 747--766.

\bibitem{Yukalov_28} 
V.I. Yukalov, 
Equilibration of quasi-isolated quantum systems, 
Phys. Lett. A 376 (2012) 550--554.

\bibitem{Yukalov_29}
V.I. Yukalov, 
Decoherence and equilibration under nondestructive measurements, 
Ann. Phys. (N.Y.) 327 (2012) 253--263. 

\bibitem{Kullback_30}
S. Kullback, R.A. Leibler, 
On information and sufficiency, 
Ann. Math. Stat. 22 (1951) 79--86.

\bibitem{Kullback_31}
S. Kullback, 
Information Theory and Statistics, 
Wiley, New York, 1959.

\bibitem{Martin_32}
E.D. Martin, 
The Behavior of Crowds: A Psychological Study, 
Harper, New York, 1920.
 
\bibitem{Sherif_33}
M. Sherif, 
The Psychology of Social Norms,
Harper Collins, New York, 1936.

\bibitem{Smelser_34}
N.J. Smelser, 
Theory of Collective Behavior,
Free Press, Glencoe, 1963.

\bibitem{Merton_35}
R.K. Merton, 
Social Theory and Social Structure,
Free Press, New York, 1968.

\bibitem{Turner_36}
R.H. Turner, L.M. Killian, 
Collective Behavior, 
Prentice-Hall Englewood Cliffs, 1993.

\bibitem{Hatfield_37}
E. Hatfield, J.T. Cacioppo, R.L. Rapson, 
Emotional Contagion,
Cambridge University, New York, 1994.

\bibitem{Brunnermeier_38}
M.K. Brunnermeier, 
Asset Pricing under Asymmetric Information: Bubbles, Crashes, Technical Analysis, 
and Herding, Oxford University, Oxford, 2001.

\bibitem{Sandholm_39}
W.H. Sandholm, 
Population Games and Evolutionary Dynamics, 
Massachusetts Institute of Technology, Cambridge, 2010.

\end{thebibliography}
\end{document}